%% file: ms.tex
\documentclass{emulateapj}

\def\kmsec{\mbox{km~s$^{\rm -1}$}}

\def\rpro{\mbox{$r$-process}}
\def\spro{\mbox{$s$-process}}
\def\ncap{\mbox{$n$-capture}}

\shorttitle{Detailed Abundances in a Stellar Stream}
\shortauthors{Roederer et al.}
\submitted{Accepted for publication in the Astrophysical Journal}

\begin{document}

\title{Characterizing the Chemistry of the Milky Way Stellar Halo: \\
Detailed Chemical Analysis of a Metal-Poor Stellar Stream }

\author{
Ian U.\ Roederer\altaffilmark{1},
Christopher Sneden\altaffilmark{1},
Ian B.\ Thompson\altaffilmark{2}, \\
George W.\ Preston\altaffilmark{2}, and
Stephen A.\ Shectman\altaffilmark{2}
}

\altaffiltext{1}{Department of Astronomy, University of Texas at Austin,
1 University Station, C1400, Austin, TX 78712-0259; 
iur@astro.as.utexas.edu}

\altaffiltext{2}{Carnegie Observatories, 
813 Santa Barbara St., Pasadena, CA 91101}

\begin{abstract}
We present the results of a detailed abundance analysis of 
one of the confirmed building blocks of the Milky Way stellar halo, 
a kinematically-coherent metal-poor stellar stream.
We have obtained high resolution and high S/N spectra of
12 probable stream members using the
MIKE spectrograph on the Magellan-Clay Telescope at Las Campanas Observatory
and the 2dCoude spectrograph on the Smith Telescope at McDonald Observatory.
We have derived abundances or upper limits for 51 species 
of 46 elements in each of these stars.
The stream members show a range of metallicity 
($-$3.4~$<$~[Fe/H]~$< -$1.5) but are otherwise chemically homogeneous,
with the same star-to-star dispersion in [X/Fe] as the rest of the halo.
This implies that, in principle, a significant fraction of the
Milky Way stellar halo could 
have formed from accreted systems like the stream.
The stream stars show minimal evolution
in the $\alpha$ or Fe-group elements over the range of metallicity.
This stream is enriched with material produced by the
main and weak components of the rapid neutron-capture process
and shows no evidence for enrichment by 
the slow neutron-capture process.
\end{abstract}


\section{Introduction}
\label{intro}

It has been a decade since \citet{helmi99b} first reported the detection
of a group of low metallicity stars whose angular momentum components
clumped together far more than would be expected for a random distribution 
of halo stars.  
This ``stellar stream'' passes through the Solar neighborhood, and 
Helmi et al.\ proposed, on the basis of models of satellite
disruption by the Milky Way, that this stream may have originated
from the disruption of a former Milky Way satellite galaxy perhaps similar 
the Fornax dwarf spheroidal (dSph) system.
The stars in this stream are scattered all around the sky and have no clear 
spatial structure, yet they have remained clumped together in velocity space.
From the number of stars in the stream, Helmi et al.\ estimated that
$~\sim$~12\% of all metal-poor stars beyond the Solar circle may have 
originated in this disrupted system.
Although the fraction of metal-poor stars estimated to have originated in
this disrupted satellite have been revised downward as metal-poor 
stellar samples have increased in size, the presence of this particular
stellar stream has been reconfirmed
by many subsequent studies (\citealt{chiba00}, \citealt{refiorentin05},
\citealt{dettbarn07}, \citealt{kepley07}, \citealt{klement09}, 
\citealt{smith09}).

Thanks to the wealth of photometric and low-resolution spectroscopic
data generated by the SDSS \citep{york00},
SEGUE \citep{yanny09}, and 2MASS \citep{skrutskie06},
a host of faint, metal-poor
stellar streams have since been discovered as stellar or 
velocity overdensities in the halo of the Milky Way.
Some of these streams are tidal debris 
associated with the disruption of the Sagittarius (Sgr) dSph 
(e.g., \citealt{vivas01,newberg02,majewski03,martinezdelgado04,belokurov06b})
or the Bootes~III dSph \citep{carlin09,grillmair09}.
Some may be the remnants of globular clusters partially or 
completely disrupted by the Milky Way 
\citep{grillmair09,newberg09}.
Others are associated with the newly-discovered ultra-faint dwarf 
galaxies (uFd; e.g., the Orphan Stream is likely affiliated with
the Ursa Major~II~uFd and several globular clusters: 
\citealt{grillmair06c,zucker06,belokurov07b}).
Still others have no known progenitor systems (e.g., \citealt{klement09}).
Some Milky Way globular clusters also show tidal tails (e.g., 
\citealt{grillmair95,odenkirchen01,grillmair06a,belokurov06a}) and 
multiple stellar populations 
(e.g., \citealt{lee99,piotto07,milone08,dacosta09,han09}).
It is possible that
some globular clusters, dwarf galaxies, stellar streams, and 
stellar overdensities originated from more complex systems
(e.g., \citealt{lyndenbell95}, \citealt{lee07}).
This hypothesis is also supported by the
comparison of simulated halo substructure to that observed 
in large surveys (e.g., \citealt{bell08,starkenburg09}), which
find that the bulk of the outer---presumably 
accreted---stellar halo can be accounted for by the 
disruption and accretion of satellites.

Because many of these structures lie at least a few tens of kpc from
the Sun, it has been virtually impossible to 
obtain detailed chemical abundances from high resolution 
spectra of individual stars in these systems.
It is more common for indirect metallicity estimates to be made 
by comparing broadband photometry with isochrones or ridge lines 
from Milky Way globular cluster fiducials.
Spectroscopically-derived abundance measurements have only been made
in several of the Sgr debris streams for a limited number
of elements (Fe, Mg, Ca, Ti, Y, and La: \citealt{monaco07,chou07,chou10}).

In contrast, globular clusters
have been popular targets for abundance studies for decades
(e.g., \citealt{gratton04} and references therein).
In fact, the globular cluster abundance literature is extensive
enough to easily identify ``outlier'' clusters whose compositions 
are distinctly different from other clusters with similar metallicities.
For example, M54 \citep{brown99},
Pal~12 \citep{cohen04a}, and Ter~7 \citep{sbordone07}
have [$\alpha$/Fe] ratios a factor of 2--3 lower than field halo stars
at the mean cluster metallicities; however, 
when their [$\alpha$/Fe] ratios are compared with those in the core
of Sgr the chemical resemblance is unmistakable.
(See, e.g., Figure~3 of \citealt{sbordone07}.)
For these clusters, the combination of kinematic and chemical information
has established an unambiguous association with Sgr.

In this spirit, we have examined the abundances of 51 species of 
46 elements in 12 candidate members
of the stellar stream identified by \citet{helmi99b}.
These stars are in the Solar neighborhood (distances $\lesssim$~2~kpc)
and are therefore bright, so it is relatively easy
to obtain high resolution spectra at high signal-to-noise (S/N) ratios
over the entire visible spectral range.
High resolution abundance studies date back
more than 30 years for two of them,
\mbox{HD~128279} \citep{spite75} and
\mbox{HD~175305} \citep{wallerstein79},
where they have traditionally been classified as members of 
``the halo.''
Now, armed with a fuller knowledge of their kinematic properties,
we are fortunate to have a more precise context in which to interpret 
their chemical enrichment patterns.

Sections~\ref{obs} and \ref{membership} describe our observations
and confirmation of the membership of individual stars.
We perform a standard abundance analysis of these stars
(Section~\ref{abund}),
with particular emphasis on deriving reliable measures of
the star-to-star abundance dispersion
(Section~\ref{smallnumber}),
and present our results in Section~\ref{results}.
The implications for chemical enrichment of the stream and 
Galactic halo, as well as plausible scenarios for the origin
of the stream, are discussed in Section~\ref{discussion}.
Our findings are summarized in Section~\ref{conclusions}.

\section{Observations}
\label{obs}

Six candidate stream members were observed with the Robert G.\ Tull
Cross-Dispersed Echelle Spectrometer
(2dCoude; \citealt{tull95}), located on the 2.7~m Harlan J.\ Smith
Telescope at McDonald Observatory.
These spectra were taken with the 2.4'' by 8.0'' slit, yielding 
a resolving power $R \equiv \lambda/\Delta\lambda \sim$~33,000.
This setup delivers complete wavelength coverage from 3700--5700\AA,
with small gaps between the echelle orders further to the red.
For our abundance analysis we only use the spectra blueward of 8000\AA.

An additional six candidate stream members were observed with the 
Magellan Inamori Kyocera Echelle (MIKE)
spectrograph \citep{bernstein03}, located on the 
6.5~m Magellan-Clay Telescope at Las Campanas Observatory.
The MIKE spectra were taken with the 0.7'' wide slit, yielding
a resolving power of $R \sim$~41,000 in the blue 
and $R \sim$~32,000 in the red.
This setup provides complete wavelength coverage from approximately
3350--9150\AA.

For data obtained with 2dCoude,
reduction, extraction, and wavelength calibration 
(derived from ThAr exposures taken before or after each stellar exposure)
were accomplished using the REDUCE software package \citep{piskunov02}.
This package is optimized for automatic reduction of data obtained with 
cross-dispersed echelle spectrographs.
For the data obtained with MIKE, reduction,
extraction, and wavelength calibration were performed using 
the current version of the MIKE data reduction pipeline
written by D.\ Kelson (see also \citealt{kelson03}).
Observations were broken into several subexposures
with exposure times typically not longer than 30~m.
Coaddition and continuum normalization were performed within the 
IRAF environment.\footnote{
IRAF is distributed by the National Optical Astronomy Observatories,
which are operated by the Association of Universities for Research
in Astronomy, Inc., under cooperative agreement with the National
Science Foundation.}

In Table~\ref{obstab} we present a record of all observations of 
the candidate members of the stream.
S/N estimates, listed in Table~\ref{rvtab}, are based on
Poisson statistics for the number of photons collected 
at several reference wavelengths
once all observations of a given target have been coadded together.
To measure the radial velocity (RV) of each of our target stars,
we cross correlate our spectra against standard template stars
using the \textit{fxcor} task in IRAF.
The RV with respect to the ThAr lamp
is found by cross correlating the echelle
order containing the Mg~\textsc{i} \textit{b} lines.
We also cross correlate the echelle
order containing the telluric O$_{2}$ B band near 6900\AA\ 
(using empirical O$_{2}$ wavelengths from \citealt{griffin73})
with a template to identify any velocity shifts resulting from 
thermal and mechanical motions in the spectrographs.
The standard deviation of these corrections is 0.4~\kmsec,
which is consistent with no shift.\footnote{
We estimate the standard deviation of the telluric zeropoint 
based on 320 individual MIKE spectra
collected by us to be published elsewhere.
We estimate the standard deviation of the total uncertainty 
based on comparison of 153 individual spectra with
independent measurements from the literature for 
RV-constant stars.
We suspect that the standard deviation of the telluric zeropoint
may be smaller, at least in part, 
due to the higher S/N and higher telluric line density
in this order than in the order containing the 
Mg~\textsc{i} \textit{b} lines.}
Velocity corrections to the Heliocentric rest frame are computed 
using the IRAF \textit{rvcorrect} task.
Heliocentric RV measurements for each observation of each star are
listed in Table~\ref{obstab}.
We estimate that this method yields a total uncertainty of 0.8~\kmsec\ 
per observation.
The mean RV derived for each target is listed in Table~\ref{rvtab}.

\section{Stream Membership}
\label{membership}

\citet{helmi99b} and subsequent investigators have identified
the presence of a stream by virtue of stellar kinematics.
Membership is always defined in terms of probabilities, and
investigators searching for the presence of streams in large
datasets are primarily concerned with the statistical detection
of a stream rather than the precise identification of which
stars are members and which are not.
Our target list was compiled from the
tables of \citet{chiba00}, \citet{refiorentin05}, and 
\citet{kepley07}.
A handful of RR~Lyrae stars have also been identified as candidate
members, which we have not observed due to the limited elements 
with detectable transitions and the difficult nature
of deriving abundances that can be compared meaningfully with abundances
from ordinary giant stars.
We adopt a strict kinematic
definition for membership as described below.

\subsection{A Kinematic Definition of Membership}

\begin{figure}
\includegraphics[angle=0,width=3.4in]{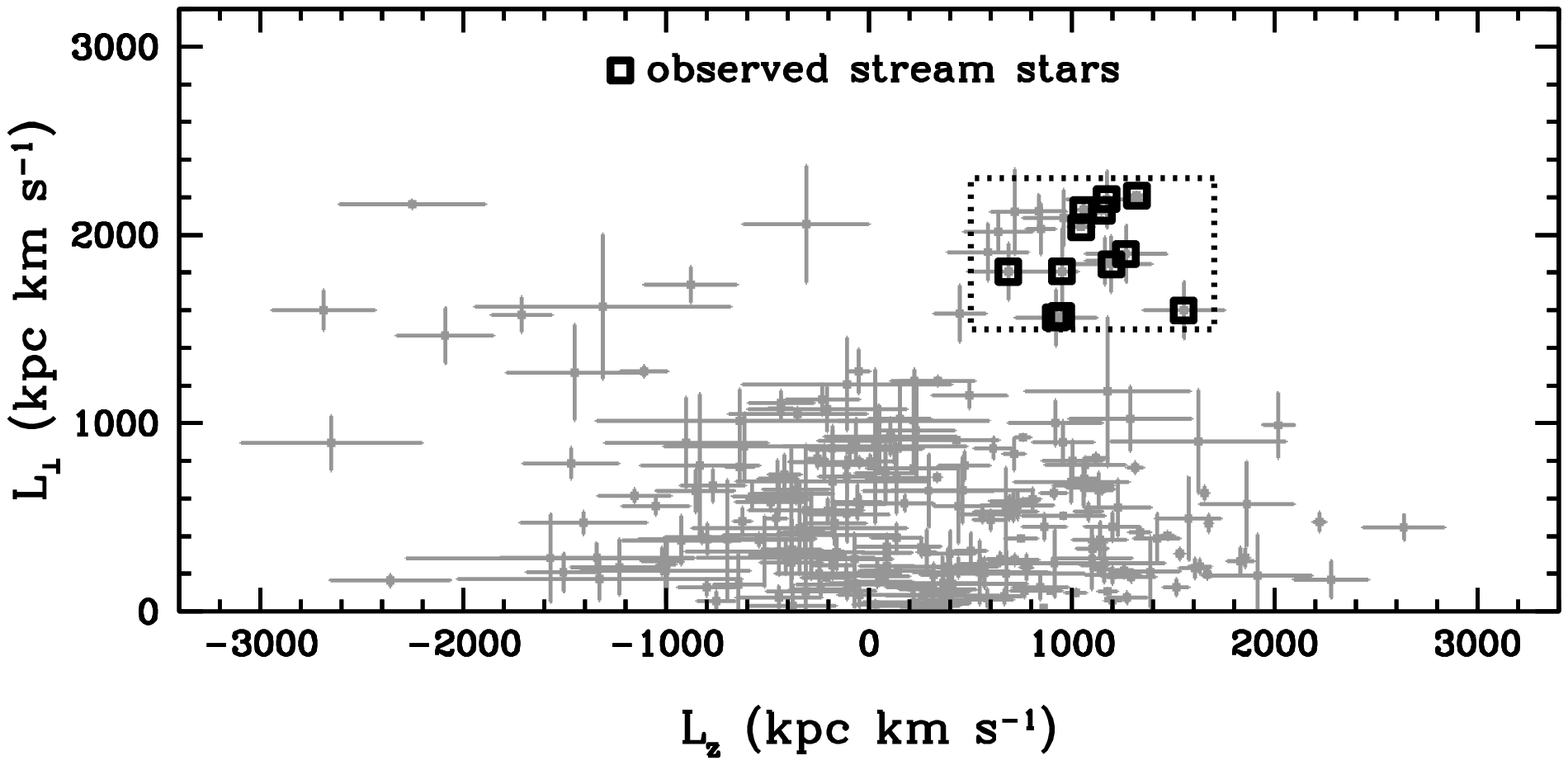}
\caption{
\label{angmomplot}
Plot of the angular momentum components $L_{z}$ versus $L_{\perp}$.
The data are mostly taken from \citet{morrison09} and supplemented
with data from \citet{chiba00} and \citet{refiorentin05} for some
of the stream stars.
The dotted box indicates the approximate range of angular momentum 
components of the stream members.
The compiled data displayed here are intended to be representative
of the distribution of angular momenta for stars in the disk and halo
to illustrate the relative range of angular momenta occupied by
the stream candidates. 
The unobserved stars in the box are RR~Lyr stars or stars too
faint to reasonably observe for our program.
}
\end{figure}

In Figure~\ref{angmomplot} we show the angular momentum components
for the stream candidates compared with other halo stars;
the data have been taken from \citet{morrison09} and supplemented
by \citet{chiba00} and \citet{refiorentin05}.
The stream stars are clumped together near 
($L_{z}$, $L_{\perp}$)~$=$~(1000, 2000)~kpc~km~s$^{-1}$.
In this representation, $L_{z} =$~0~kpc~km~s$^{-1}$ corresponds to
no net rotation about the Galactic center, and increasing $L_{\perp}$
corresponds to orbits increasingly tilted out of the plane of the disk.
For reference, the Sun and other stars of the thin disk have 
($L_{z}$, $L_{\perp}$)~$\approx$~(1800, 0)~kpc~km~s$^{-1}$.
The candidate stream members have prograde, eccentric orbits that take
them well above the Galactic plane:
minimum Galactocentric distance ($R_{\rm peri}$) of 7~kpc, 
maximum Galactocentric distance ($R_{\rm apo}$) of 16~kpc, 
and maximum distance ($|Z|_{\rm max}$) of 13~kpc above the Galactic plane
\citep{helmi99b}.

To compute the orbit of an individual star, one needs to know its
coordinates, distance, radial velocity, and proper motion;
the position and space motion of the star can then be integrated
over time in a model of the Galactic potential.
Our study remeasures one of these quantities, the radial velocity,
so we have checked our RV measurements against those employed by previous
investigators to derive the kinematic properties of the stars.
In some cases, the previous RV measurements were made from 
medium resolution spectra.
If our high resolution 
RV measurements differ from the previous measurement,
the original set of kinematic properties is suspect;
fortunately, the difference can be quantified.
It is also important that the stars are not in binary systems
or, if they are, that the RV used to compute the kinematic properties
is the systemic velocity.
Comparing with previous high resolution and multi-epoch RV measurements by
\citet{latham91}, \citet{carney03,carney08}, and \citet{zhang09},
we confirm that this is the case.
We retain all 12 candidates, listed in Table~\ref{rvtab}, 
as probable stream members. 
Below, we discuss a few candidates in more detail.

\subsection{Comments on Individual Candidates}

\citet{beers92} initially measured the RV of \mbox{CS~22948--093} 
as 395~\kmsec, which was the value use to compute the original 
kinematics for this star.
This star is RV variable over the 3-year baseline of our measurements,
and the RV differs by approximately 30~\kmsec\ from the \citet{beers92}.
From the definitions of $L_{z}$ and $L_{\perp}$ and the
measured or derived stellar quantities from \citet{beers00}, 
we can quantify the
effect this RV difference has on the derived angular momenta,
($\Delta L_{z}$, $\Delta L_{\perp}$)~$\approx$~(0, 200)~kpc~km~s$^{-1}$.
This is not sufficient to move \mbox{CS~22948--093} from the main
locus of probable stream members in Figure~\ref{angmomplot}, 
so we retain it as a member.

Our abundance analysis of \mbox{CS~29513--032} indicates that it 
has been polluted by material from a companion star that passed through
the asymptotic giant branch (AGB) phase of evolution.
(See Appendix~\ref{cs29513m032text}.)
This implies that \mbox{CS~29513--032} is in a binary
or multiple star system, but we have detected 
no RV variations over a span of 3~months.
Assuming that the star is in a binary system and that 
the systemic velocity is as much as 20~\kmsec\ 
different from the measured RV would imply
($\Delta L_{z}$, $\Delta L_{\perp}$)~$\approx$~(10, 160)~kpc~km~s$^{-1}$.
This is not sufficient to move \mbox{CS~29513--032} 
much farther from the other probable members, so
we also include this star as a member.

We made three observations of \mbox{CS~29513--031} 
with a time interval of more than three years.
\citet{beers92} measured a RV of 295~\kmsec\ for this star.
Our measurements indicate that this star is also RV variable, but 
the systemic velocity is not likely to be more than 
$\approx$~10~\kmsec\
different from the \citet{beers92} RV.
If the systemic velocity is 10~\kmsec\ different than the mean
of the measured RV, 
($\Delta L_{z}$, $\Delta L_{\perp}$)~$\approx$~(300, 80)~kpc~km~s$^{-1}$.
\mbox{CS~29513--031} is already located near the locus of 
probable members, and these uncertainties are small relative to the
range of angular momenta for probable members, thus
we also retain this star as a member.

\section{Abundance Analysis}
\label{abund}

\subsection{Linelist and Equivalent Width Measurement}

We measure equivalent widths from our spectra using a 
semi-automatic routine that fits Voigt absorption line profiles
to continuum-normalized spectra.
Each measurement must be visually inspected and approved by the user.
When possible, we use a single source for all of the log($gf$) values
for a given species.  
References for our log($gf$) values are given in Table~\ref{loggftab}.
Our equivalent width measurements and atomic data are presented
in Table~\ref{ewtab}.

The Na~\textsc{i} lines at 5889 and 5895~\AA\ are often contaminated by
telluric absorption features, and we have only measured equivalent
widths for these lines when they appear to be velocity-shifted
away from atmospheric components.
We assess this by comparing the atmospheric transmission spectrum 
to the observed stellar spectrum.
Additionally, the Na stellar lines may be blended with absorption or
emission components from Na in the interstellar medium (ISM), and we only 
measure equivalent widths for these lines when no interstellar components
are detected (e.g., as line asymmetries).
We adopt the same technique to avoid telluric contamination to all lines
redward of 5660\AA, especially the high-excitation lines of 
Na~\textsc{i} and Si~\textsc{i}, which are often weak and might easily
be mistaken for telluric absorption, 
and the K~\textsc{i} resonance lines at 7664 and 7698\AA.

\subsection{Derivation of the Model Atmosphere Parameters}

We perform the abundance analysis using the latest version (2009)
of the spectral analysis code MOOG \citep{sneden73}.
This version of MOOG incorporates the contribution of electron
scattering in the near-UV continuum as true continuous opacity rather
than extra absorption. 
(See \citealt{sobeck10} for additional details.)
Throughout our analysis we assume that all lines are formed under
conditions of local thermodynamic equilibrium (LTE) 
in a one dimensional, plane parallel atmosphere.

Model atmospheres are interpolated from the grid of \citet{castelli03},
generated using the $\alpha$-enhanced opacity distribution functions
assuming no convective overshooting.
Atmospheric parameters are derived by spectroscopic means only.
The effective temperature and microturbulence are determined by
demanding that the 
derived Fe~\textsc{i} abundances exhibit no trend with excitation potential 
or reduced equivalent width (i.e., equivalent width divided by wavelength).
The resulting temperature and microturbulence generally satisfy these
two criteria for the 
Fe~\textsc{ii} and Ti~\textsc{i} and \textsc{ii} abundances,
though there are typically fewer lines of these species
(8--11, 7--16, and 12--20 lines, respectively, compared with
80--120 lines for Fe~\textsc{i}).
These three species have a much 
more limited range of excitation potentials (typically $\sim$~1~eV, 
compared with $\sim$~4.5~eV for Fe~\textsc{i}).
The model surface gravity is determined by requiring that the
Fe~\textsc{i} and \textsc{ii} abundances agree within 0.1~dex 
(roughly the standard deviation of each).
The model metallicities are set to the Fe abundance.
The $\alpha$ elements
are enhanced by a factor of 2--3 relative to Fe in our stars
(Section~\ref{results}).
These elements are among the major electron donors to the 
H$^{-}$ ion that dominates the
continuous opacity over the visible spectral range, 
justifying our use of the $\alpha$-enhanced grid of atmospheres.

\begin{figure}
\includegraphics[angle=0,width=3.4in]{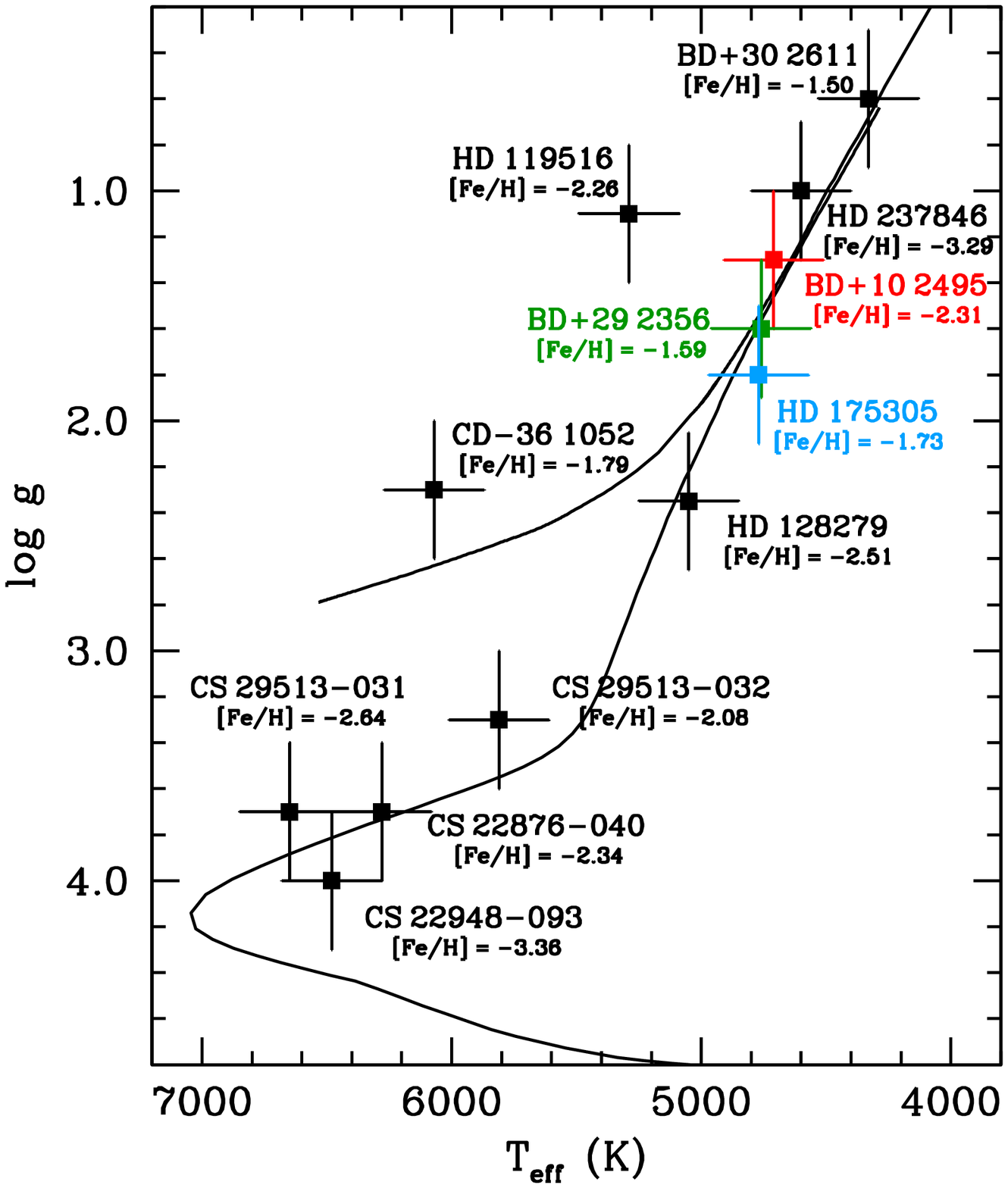}
\caption{
\label{cmd}
Plot of the evolutionary states of the stream candidates.
Our spectroscopically derived temperatures and gravities are shown, 
along with uncertainties of $\pm$~200~K and $\pm$~0.3~dex.
Three stars along the RGB are color-coded to match
the spectra displayed in Figure~\ref{specplot}:
\mbox{BD$+$10~2356},
\mbox{BD$+$29~2495}, and
\mbox{HD~175305}.
For comparison, a Y$^{2}$ $\alpha$-enhanced ([$\alpha$/Fe]~$=+0.4$)
8~Gyr isochrone 
\citep{demarque04} and a synthetic horizontal branch track 
for 0.8~M$_{\odot}$ \citep{cassisi04} are shown, 
each computed for [Fe/H]~$=-2.0$.
Stream members are indicated as filled squares, and
rejected candidates are indicated as ``X''s.
}
\end{figure}

Figure~\ref{cmd} displays the evolutionary states of these stars, 
and Table~\ref{modeltab} lists our derived model parameters.
Six of them are found along the red giant branch (RGB),
two are near the red horizontal branch (RHB),
and four are on the subgiant branch (SGB) or main sequence turn-off.  
This range of evolutionary states will complicate inter-comparison of the
absolute abundances; element-to-element ratios, however, should be 
more reliable, especially if the abundances are derived from lines
of similar strength and thus are formed at similar levels of the atmosphere.

In Table~\ref{teffcomp} we compare our spectroscopic temperatures 
to those derived from the infrared flux method \citep{alonso99a,alonso99b}
for five of the six stars on the RGB.
The stars on the SGB and RHB are beyond the range of the calibrations.
Five of the giants (\mbox{BD$+$10~2495}, 
\mbox{BD$+$30~2611}, \mbox{HD~128279}, \mbox{HD~175305}, and \mbox{HD~237846})
were included among the metal-poor calibration stars used by 
\citet{alonso99a}, and we report their temperature estimates in 
Table~\ref{teffcomp}.\footnote{
The only published $V$ magnitude for the remaining giant, 
\mbox{BD$+$29~2356}, dates from
a photoelectric measurement by \citet{harris64}, which we
disregard in the interest of self-consistency with the \citet{alonso99a}
calibrations.}
The spectroscopic and photometric $T_{\rm eff}$ estimates agree within
the uncertainties for the coolest star in the sample, \mbox{BD$+$30~2611}, 
but for three of the other giants our spectroscopic estimates 
are cooler than the photometric ones by $\approx$~250~K
and the fourth is cooler by $\approx$~380~K.
This star, \mbox{HD~237846}, was also analyzed by
\citet{zhang09}, who derived a spectroscopic temperature
of 4725~$\pm$~60~K, which is only different from our
estimate by 125~K.

\begin{figure}
\includegraphics[angle=0,width=3.4in]{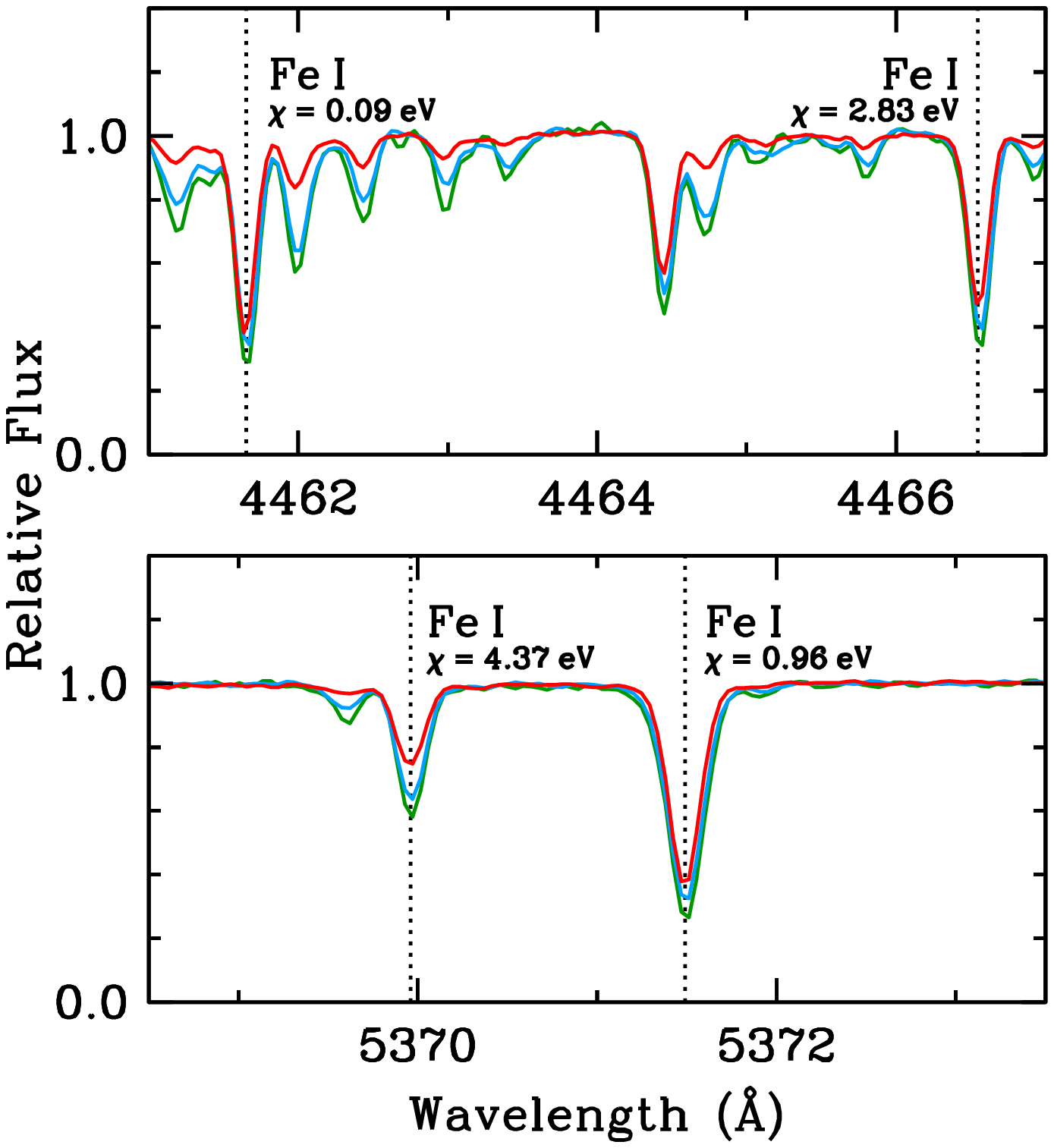}
\caption{
\label{specplot}
Comparison of observed spectra for two pairs of Fe~\textsc{i} lines 
with differing excitation potentials ($\chi$).
Three stars along the RGB with very similar $T_{\rm eff}$ are shown:
\mbox{BD$+$10~2356}, red ($T_{\rm eff}=$4710~K, [Fe/H]$= -$2.31);
\mbox{BD$+$29~2495}, green (4760~K, [Fe/H]$= -$1.59); and
\mbox{HD~175305}, blue (4770~K, $-$1.73).
The colors also correspond to those used in Figure~\ref{cmd}.
}
\end{figure}

In Figure~\ref{specplot}, we compare the spectra of three
of the stars on the RGB with very similar temperatures
but different metallicities.
Two pairs of Fe~\textsc{i} lines are shown, each 
with different excitation potentials.
(By choosing line pairs at the same wavelength, we 
can compare line strength without the continuous opacity
changing appreciably.)
Our spectroscopic analysis shows that \mbox{BD$+$10~2495},
\mbox{BD$+$29~2356}, and \mbox{HD~175305} all have temperatures
within 60~K of one another, so to first order the largest difference
in the line strengths in these stars is the Fe abundance;
Figure~\ref{specplot} indicates that this is the case.
Thus in a relative sense our temperatures are reasonable.

It is more difficult to assess the absolute temperature scale,
but in a differential abundance analysis this effect is minimized
as much as possible.
In light of the discrepancies between the photometric and
spectroscopic temperature scales, 
we adopt uncertainties on $T_{\rm eff}$, log~$g$, and
$v_{t}$ of 200~K, 0.3~dex, and 0.3~\kmsec.
We adopt spectroscopic methods to estimate atmospheric parameters to avoid
reliance on photometry (which may originate from a variety of 
sources, especially for bright stars) 
or models of the flux distribution (which must reproduce the 
energy distribution in stars of a variety of evolutionary states
and compositions).
Spectroscopically determined atmospheric parameters are certainly
not without their own limitations, 
but the fact that most of the stars in Figure~\ref{cmd} lie along
theoretical isochrones or HB tracks is reassuring in this regard.

\subsection{Derivation of Abundances}

For species with equivalent width measurements, we derive the
abundance by forcing the individual line abundances to match
the equivalent width and averaging over all lines.
We match synthetic to observed spectra to derive abundances for
species whose lines
are broadened by hyperfine splitting, isotope shifts, or both,
as well as for lines that are often weak or blended in our spectra
(Li, Al, Sc, V, Mn, Co, Sr and all heavier elements).
The C abundance is derived from a synthesis of the CH 
$A^2\Delta - X^2\Pi$ G-band 
between 4290 and 4330~\AA, and the
N abundance is derived from a synthesis of the CN 
$B^2\Sigma - X^2\Sigma$ 
bandhead region near 3880\AA.
For the odd-$Z$ Fe-peak species, we adopt the hyperfine
structure patterns of \citet{kurucz95}.
We use an \rpro\ mix of Ba isotopes in our synthesis of the Ba~\textsc{ii}
4554\AA\ line for all stars except \mbox{CS~29513--032}, where we use
an \spro\ mix (Appendix~\ref{cs29513m032text}).
For all other Ba~\textsc{ii} lines, the isotopic mix has
no noticeable effect on the derived abundance.
We adopt the hyperfine structure patterns for the rare earth elements 
from the references summarized in \citet{lawler09}.
Table~\ref{ewtab} lists all of our equivalent width measurements or else
indicates whether a line is used to determine an abundance via
spectral synthesis or whether we compute an upper limit on the abundance
from the line.

Abundances for all 12~stars are presented in 
Tables~\ref{abundtab1}--\ref{abundtab4}.
Our main aim in this work is to compare the abundances of stars
in the stream to each other and to other metal-poor populations.
To this end, we do not apply any corrections to the abundances
from our 1D LTE analysis to account for hydrodynamical motions
or departures from LTE in the real atmospheres of these stars.
We reference abundance ratios to the Solar abundances
summarized in \citet{asplund09}.

We assume a minimum uncertainty of 0.25~dex for abundances derived
from a single spectral feature 
(set by statistical sources of error: 
our ability to resolve blending features, 
identify the continuum, measure an equivalent
width or match a synthetic spectrum, etc.).
For mean abundances derived from two lines, 
we adopt the larger of 0.15~dex or
the standard deviation of a small sample as described by \citet{keeping62}.
For mean abundances derived from more than two lines,
we adopt the larger of of 0.10~dex or the standard deviation.
We compute 3$\sigma$ abundance upper limits from the non-detection of
absorption lines according to the formula given in \citet{frebel08}, 
which was derived from \citet{bohlin83}.
These upper limits are indicated in Table~\ref{abundtab1}--\ref{abundtab4}.

\section{Systematic Abundance Trends}
\label{smallnumber}

It is important to recognize systematic differences that
can bias results or mask real trends when performing
detailed abundance comparisons.
Stars of different metallicities or evolutionary states
will naturally present different atomic transitions 
suitable for abundance analysis, and the abundances derived from
these transitions need to be checked against one another.
In this section we examine several systematic biases that 
could impact our results.

\begin{figure*}
\includegraphics[angle=0,width=7.0in]{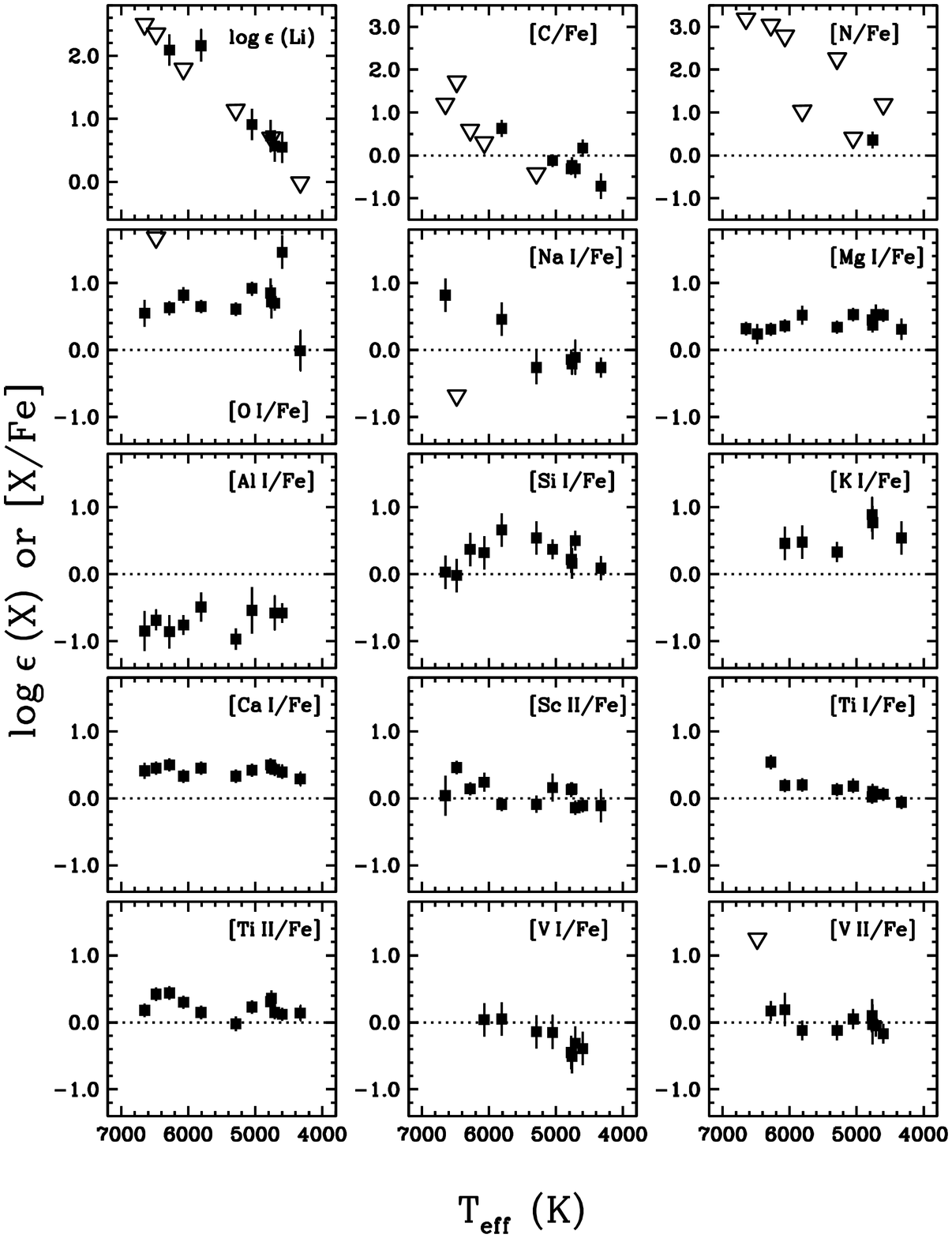}
\caption{
\label{abundteff1}
Abundances for Li~\textsc{i}--V~\textsc{ii} 
as a function of stellar effective temperature.
Squares indicate measurements in the stream members and
downward-facing triangles indicate upper limits.
The dotted lines indicate the Solar ratio.
}
\end{figure*}

\begin{figure*}
\includegraphics[angle=0,width=7.0in]{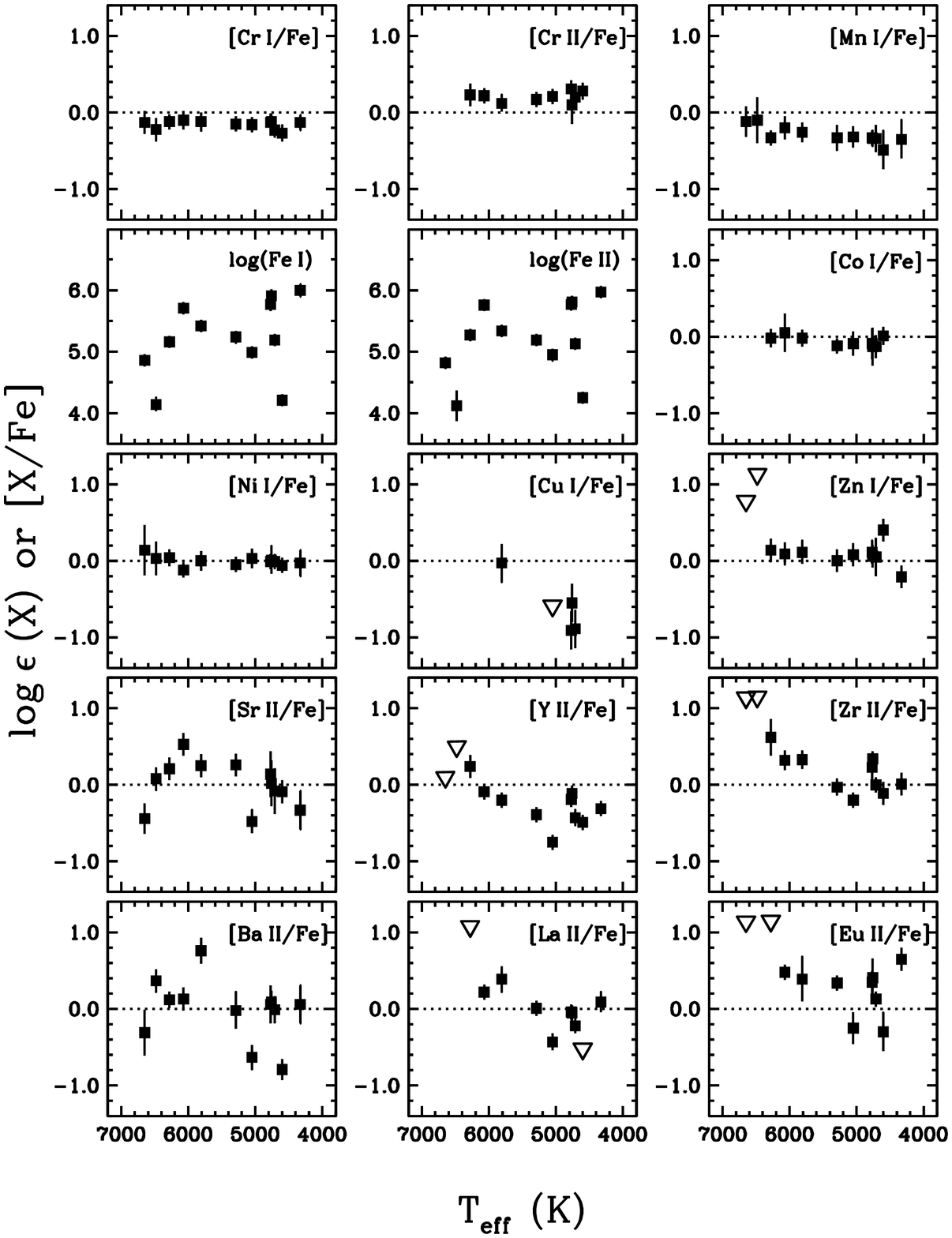}
\caption{
\label{abundteff2}
Abundances for Cr~\textsc{i}--Eu~\textsc{ii} 
as a function of stellar effective temperature.
Squares indicate measurements in the stream members and
downward-facing triangles indicate upper limits.
The dotted lines indicate the Solar ratio.
}
\end{figure*}

If the model atmosphere accurately reflects conditions in the
stellar atmosphere, the abundance of elements 
not manufactured or destroyed during normal stellar evolution
should show no dependence on $T_{\rm eff}$ when comparing 
stars with similar compositions but different temperatures.
In Figures~\ref{abundteff1} and \ref{abundteff2} we display 
the abundances of 30~species in our sample as a function of 
$T_{\rm eff}$.
The depletion of the light element Li is evident in the more
evolved (cooler) stars of the sample.
The coolest star in the sample appears to have a slightly lower
[C/Fe] ratio than the warmer stars.
The CN band was only detected in one star.
These trends are known consequences of normal stellar evolution.
Correlations with $T_{\rm eff}$ are also detected for 
[Na~\textsc{i}/Fe],
[Ti~\textsc{i}/Fe], and
[V~\textsc{i}/Fe].
These correlations will be discussed in the sections below.

\subsection{Silicon, Titanium, and Vanadium}
\label{si}

In very metal-poor (or warm) stars, the only accessible
line of Si~\textsc{i} in the visible spectral range is the 3905\AA\ line,
which may become saturated in more metal-rich (or cooler) stars;
this line has an excitation potential of 1.9~eV.
In more metal-rich stars, high-excitation (4.9--5.1~eV) Si~\textsc{i}
lines at 5665, 5701, 5708, and 5772\AA\ may be used as abundance
indicators instead.  
In our sample, we only derive an abundance from the 3905\AA\ line
in the four warmest stars,
while between 1 and 4 of the high-excitation lines are used
in the cooler stars.
The low- and high-excitation lines are not used together in any stars.
As shown in Figure~\ref{abundteff1},
the seven coolest stars ($T_{\rm eff} <$~6000~K)
with detected Si~\textsc{i} lines
all employ the high-excitation lines and show
a slope of decreasing [Si~\textsc{i}/Fe] with decreasing temperature.
This trend is in the opposite sense from what previous studies 
have uncovered.
In contrast, the four warmest stars ($T_{\rm eff} >$~6000~K)
all employ the 3905\AA\ line and may show a slope of
increasing [Si~\textsc{i}/Fe] with decreasing temperature,
which has been recognized by previous investigators
\citep{cohen04b,preston06,lai08,bonifacio09}.
Using only abundances derived from the 3905\AA\ line, 
\citet{preston06} compared the [Si/Fe] ratio in stars on the 
lower RGB to stars on the RHB (i.e., with the same temperature but different
gravities) and found no dependence on gravity, implying that this
observed difference is not a consequence of stellar evolution.
In addition to noting this trend of 
increasing [Si~\textsc{i}/Fe] with decreasing temperature,
\citet{lai08} also identified opposite trends for
[Ti~\textsc{i}/Fe] and [Ti~\textsc{ii}/Fe].

We also find a weak trend of increasing [Ti~\textsc{i}/Fe] and 
increasing [V~\textsc{i}/Fe] with increasing $T_{\rm eff}$ in our sample.
The Ti~\textsc{i} abundance is derived from $\sim$~5--15 lines, 
suggesting that this trend is not a consequence of line blending.
The V~\textsc{i} abundance is derived from a single transition, 4379.23\AA.
The direction of this trend, increasing [V~\textsc{i}/Fe] with 
increasing $T_{\rm eff}$, would imply that the blending 
feature is decreasing in line strength with increasing $T_{\rm eff}$.
No plausible blending atomic features are found in the Kurucz linelists
or the NIST database.
A $^{12}$CH transition at 4379.24\AA\ could, in principle, produce
the observed effect, although our syntheses indicate that completely
removing the contribution from CH even in the giants will increase the
V~\textsc{i} abundance by no more than 0.01--0.02~dex, far
smaller than the observed abundance change with $T_{\rm eff}$.
We have not investigated the consequences of including 3D effects
or departures from LTE in the line formation.

In summary, it it not clear what causes these trends, 
but they do not appear to be the result of any 
shortcomings unique to our analysis.
Further investigation of the 
[Si~\textsc{i}/Fe] (high-excitation lines),
[Ti~\textsc{i}/Fe], and
[V~\textsc{i}/Fe] abundance trends with $T_{\rm eff}$ is beyond
the scope of this work.
Whatever the cause, it is responsible for producing the larger 
star-to-star dispersion in these ratios than observed
for other species that show no correlation with $T_{\rm eff}$.

\subsection{Magnesium}

The Mg~\textsc{i} transitions at 3829, 5172, and 5183\AA\ have
$\chi \approx$~2.7~eV, and the Mg~\textsc{i} 
lines at 4057, 4167, 4702,
5528, and 5711\AA\ have $\chi \approx$~4.3~eV.
This fact manifests itself as a difference in line strengths.
In warm or very metal-poor stars, only the lower-excitation lines
are strong enough to be detected.
For Mg, we derive abundances from the 5172 and 5183\AA\ lines
in four stars, three of which also use several of the high-excitation lines
(\mbox{CS~22876--040}, \mbox{CS~29513--031}, and \mbox{HD~237846},
two stars on the SGB and one on the RGB).
In these three cases, the 5172 and 5183\AA\ lines yield abundances
higher than the high-excitation lines 
by 0.13, 0.29, and 0.23~dex, respectively.
\citet{cohen04b} examined this effect in seven metal-poor dwarf stars
from their sample. 
After correcting for the differences in log($gf$) values between 
the two studies, their mean offset, 0.26~dex, is very similar to ours.
The 5172 and 5183\AA\ lines also yield abundances higher than the 
3829\AA\ line by 0.32, 0.26, and 0.14~dex in 
\mbox{CS~22876--040}, \mbox{CS~22948--093}, and \mbox{HD~237846}
(again, two stars on the SGB and one on the RGB).
We find no dependence of [Mg/Fe] on $T_{\rm eff}$ in Figure~\ref{abundteff1}.

We have derived Mg~\textsc{i} abundances from at least 2 high-excitation
lines in all but one star, so 
we only adopt the Mg abundance derived from these lines.
Only in \mbox{CS~22948--093}, where no high-excitation
lines were measured, do we employ the low-excitation lines.
We omit this star when computing the [Mg/Fe] dispersion in the stream.

\subsection{Manganese}
\label{mn}

The Mn~\textsc{i} resonance triplet at 4030, 4033, and 4034\AA\ 
has an excitation potential of 0.0~eV, and
the Mn~\textsc{i} lines farther to the red
have excitation potentials ranging from 2.1--3.1~eV.
The triplet lines are the only Mn~\textsc{i}
abundance indicators in the optical regime for very metal-poor stars.
The Mn~\textsc{i} triplet is known to yield abundances
lower by 0.3--0.4~dex relative to the higher-excitation Mn~\textsc{i} lines
(e.g., \citealt{cayrel04}).
This effect is also observed in our sample, with an average difference of 
$-$0.32~dex when both the triplet and the higher-excitation lines
are detected and measurable (6~stars).
To compare relative [Mn~\textsc{i}/Fe] abundances for stars in our sample,
we correct all of the Mn~\textsc{i} triplet abundances by $+$0.3 dex.
This correction is reflected in Tables~\ref{abundtab1}--\ref{abundtab4}.
We emphasize that this is strictly an empirical correction.
In one star from our sample, \mbox{HD~128279}, we were also able to derive
an abundance of Mn~\textsc{ii} from the 3488 and 3497\AA\ lines.
We are encouraged that
the abundance derived from these lines, 
log~$\epsilon$~(Mn~\textsc{ii})~$= +2.56$~dex,
is in very good agreement with the Mn~\textsc{i} abundance derived from
the high-excitation lines, log~$\epsilon$~(Mn~\textsc{i})~$= +2.52$~dex.

\subsection{Sodium}

The Na~\textsc{i} resonance lines at 
5889 and 5895\AA\ have $\chi =$~0.0~eV, while the 5682 and 5688\AA\ lines
have $\chi \approx$~2.1~eV.
Only the 5889 and 5895\AA\ lines are detected in warm or very
metal-poor stars.
We only derive an abundance from the 5889\AA\ line in one star,
\mbox{CS~29513--031}, and we derive an abundance from the
5895\AA\ line in only one star, \mbox{CS~29513--032}.
All other Na abundances are derived from the 5682 and 5688\AA\ lines, 
which are in very good agreement with one another.
We omit \mbox{CS~29513--031} and \mbox{CS~29513--032} when 
computing the [Na/Fe] dispersion in the stream.

\subsection{Other Elements}

Abundances are derived from the resonance line pairs of 
Al~\textsc{i} (3944 and 3961\AA),
K~\textsc{i} (7664 and 7698\AA), and
Sr~\textsc{ii} (4077 and 4215\AA), as well as
the high-excitation lines of Zn~\textsc{i} (4722 and 4810\AA).
For each of these species, these lines are the only abundance 
indicators available to us.
In cases where both lines are measured, we find no systematic offsets
from one line to the next for Al, K, and Zn.  

Both Sr~\textsc{ii} resonance lines 
were measured in 10 stars, and we find that the
4077\AA\ line gives an abundance lower by 0.08 ($\sigma =$~0.04) dex
than the 4215\AA\ line.
The offset is smaller in the stars on the SGB (0.05~dex)
than stars on the RGB or RHB (0.09~dex), suggesting that 
this offset may originate, at least in part, in modeling the
line formation, as opposed to a systematic offset in the log($gf$) values.
These Sr lines are often very strong and are blended at this
metallicity range, with observational uncertainties often 0.2--0.3~dex,
so this finding should be viewed with necessary caution.
We make no correction to the Sr~\textsc{ii} abundances here, but 
this possible systematic effect should be investigated further
in larger samples at low metallicity where the blending is less severe
and the Sr lines less saturated.

\section{Abundances in the Stream Stars}
\label{results}

\begin{figure*}
\includegraphics[angle=0,width=7.0in]{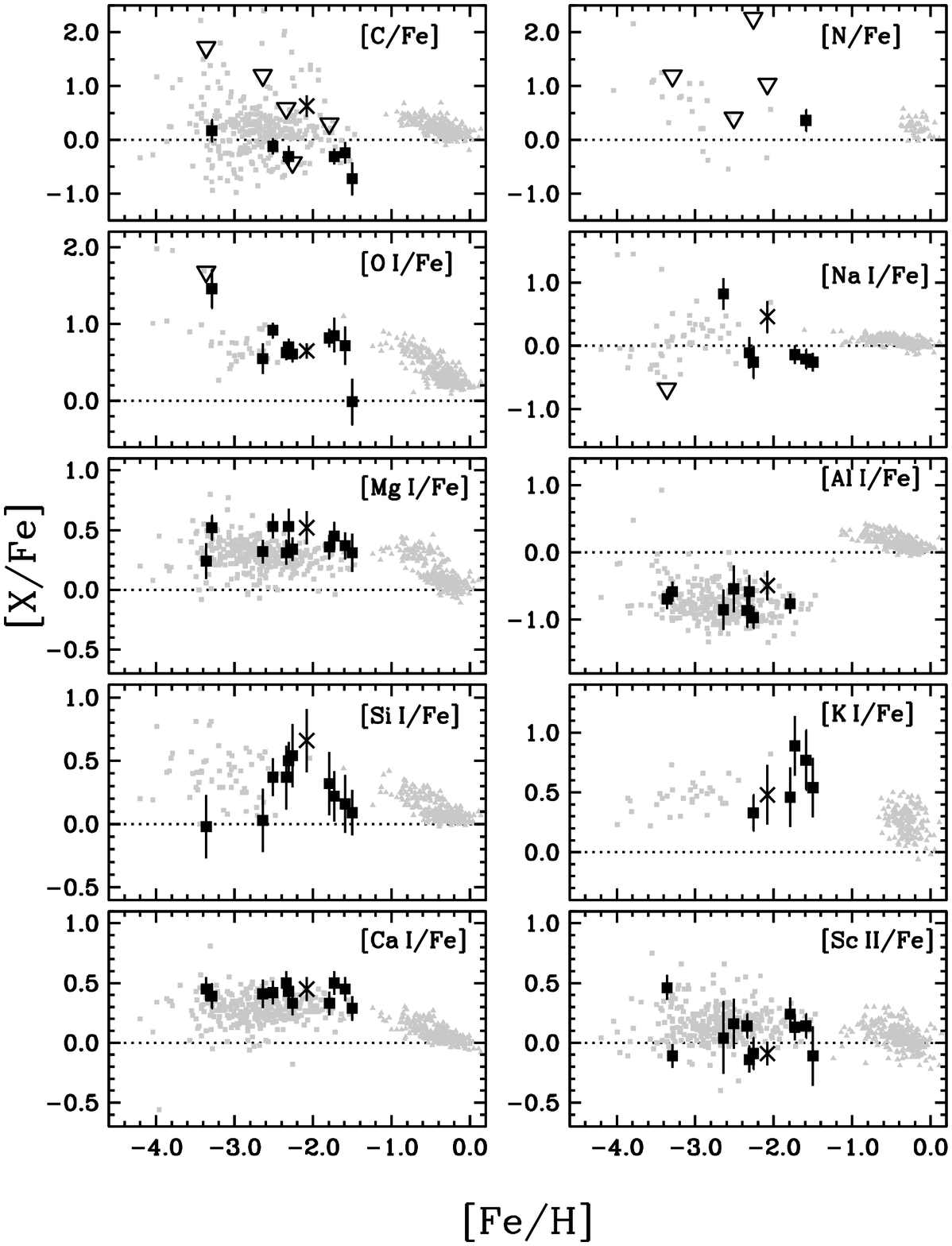}
\caption{
\label{abundplot1}
Abundance ratios [X/Fe] for C--Sc~\textsc{ii} as a function of metallicity.
Large filled squares indicate measurements in the stream members and
downward-facing triangles indicate upper limits.
\mbox{CS~29513--032} is indicated by the ``X.''
Gray triangles indicate thin and thick disc stars \citep{reddy03,reddy06}.
Gray squares indicate field halo stars
\citep{cayrel04,barklem05,cohen08,lai08}.
The Solar ratio is indicated in each panel by the dotted line.
}
\end{figure*}

\begin{figure*}
\includegraphics[angle=0,width=7.0in]{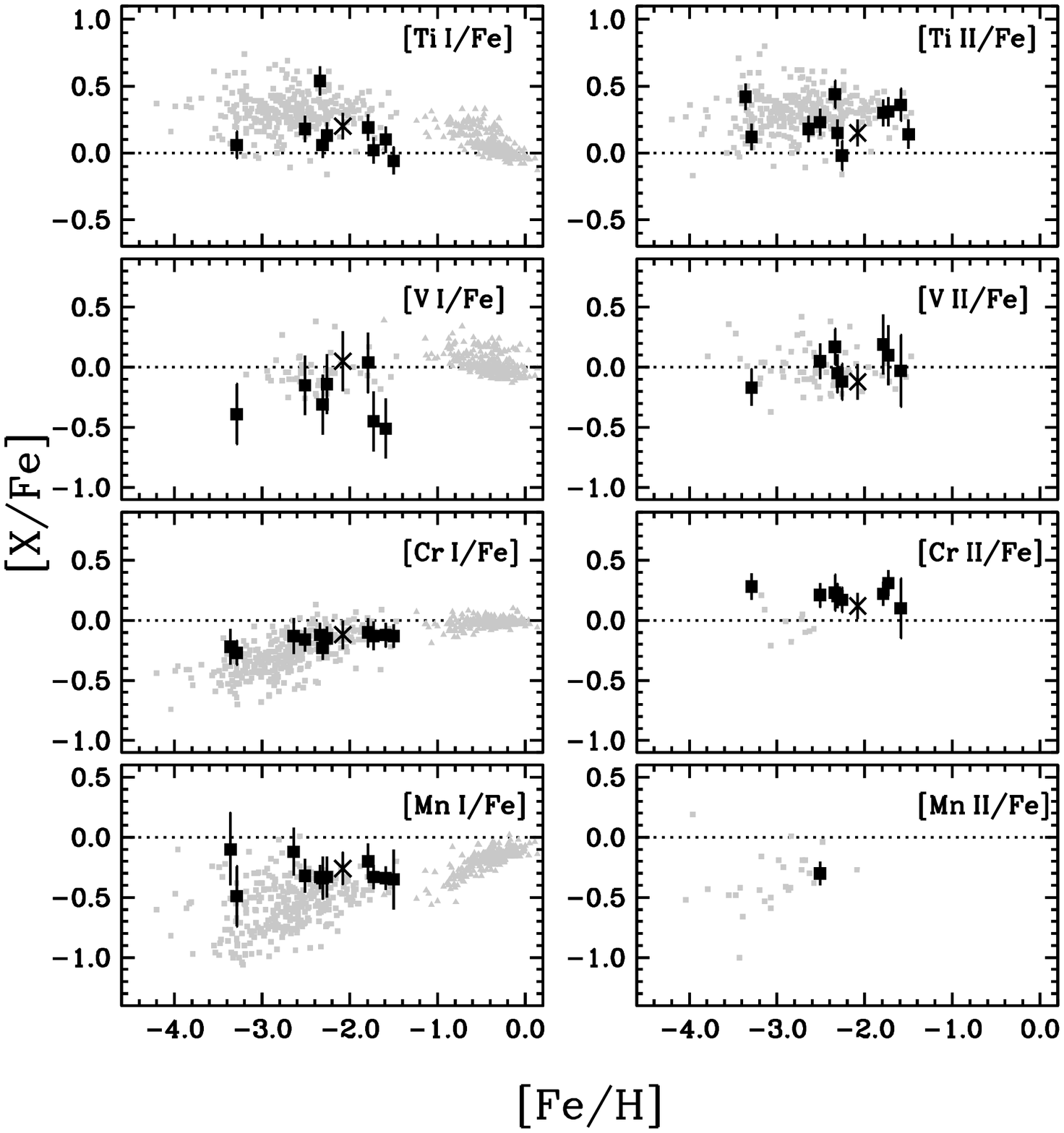}
\caption{
\label{abundplot2}
Abundance ratios [X/Fe] for Ti~\textsc{i}--Mn~\textsc{ii} 
as a function of metallicity.
Symbols are the same as in Figure~\ref{abundplot1}.
}
\end{figure*}

\begin{figure*}
\includegraphics[angle=0,width=7.0in]{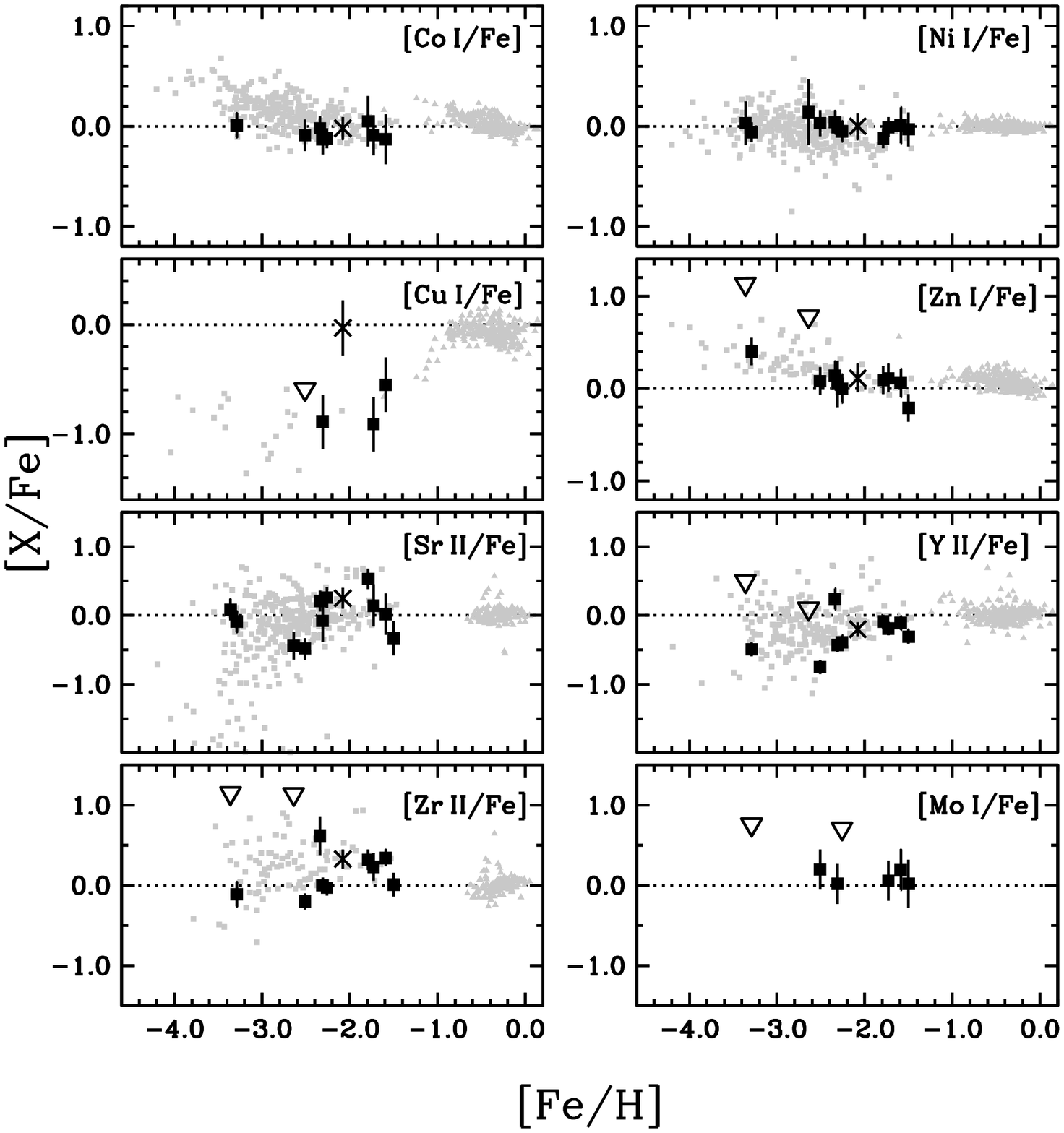}
\caption{
\label{abundplot3}
Abundance ratios [X/Fe] for Co~\textsc{i}--Mo~\textsc{i}
as a function of metallicity.
Symbols are the same as in Figure~\ref{abundplot1} and include
abundances from \citet{francois07}.
}
\end{figure*}

\begin{figure*}
\includegraphics[angle=0,width=7.0in]{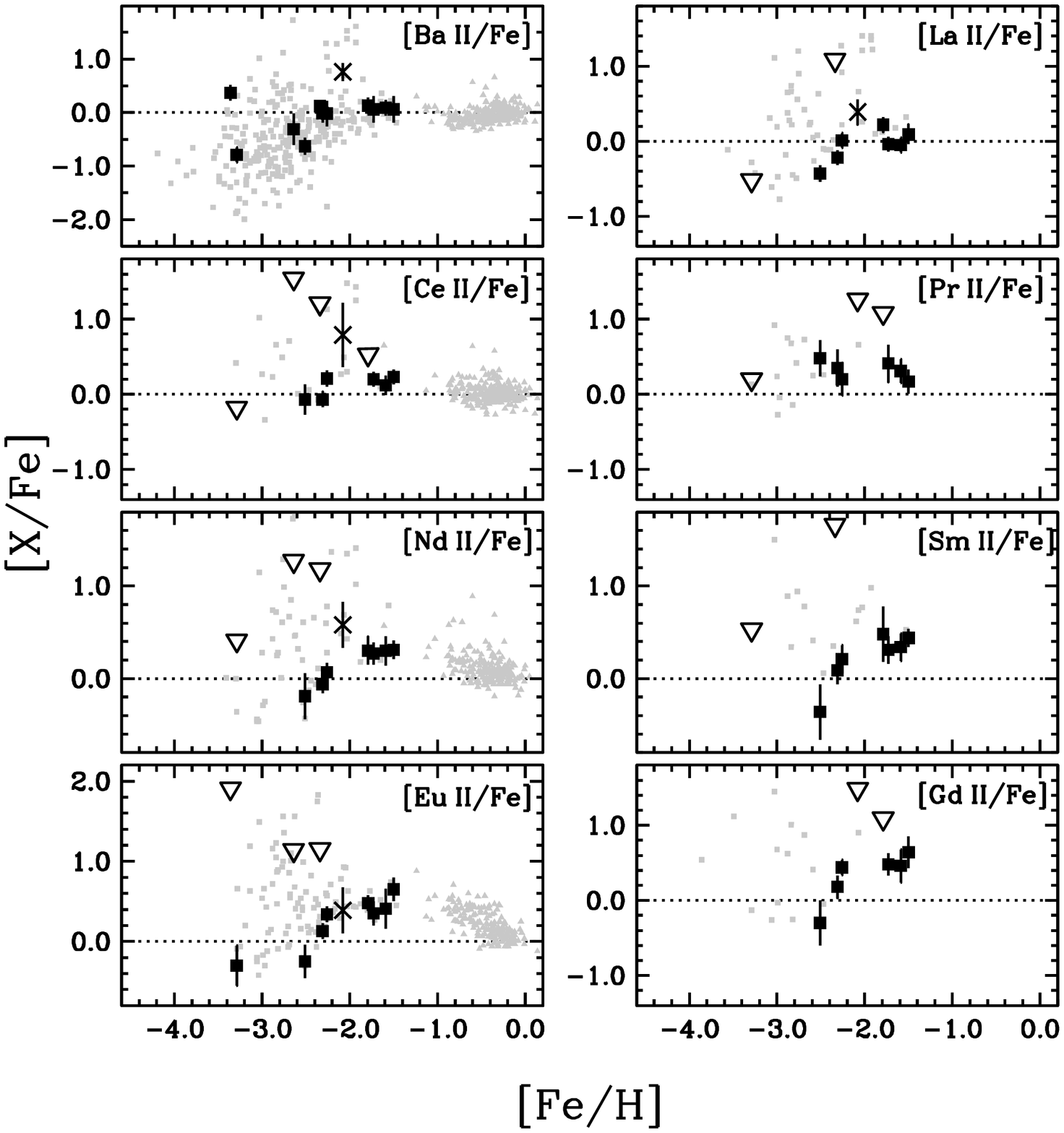}
\caption{
\label{abundplot4}
Abundance ratios [X/Fe] for Ba~\textsc{ii}--Gd~\textsc{ii}
as a function of metallicity.
Symbols are the same as in Figure~\ref{abundplot1}.
}
\end{figure*}

\begin{figure*}
\includegraphics[angle=0,width=7.0in]{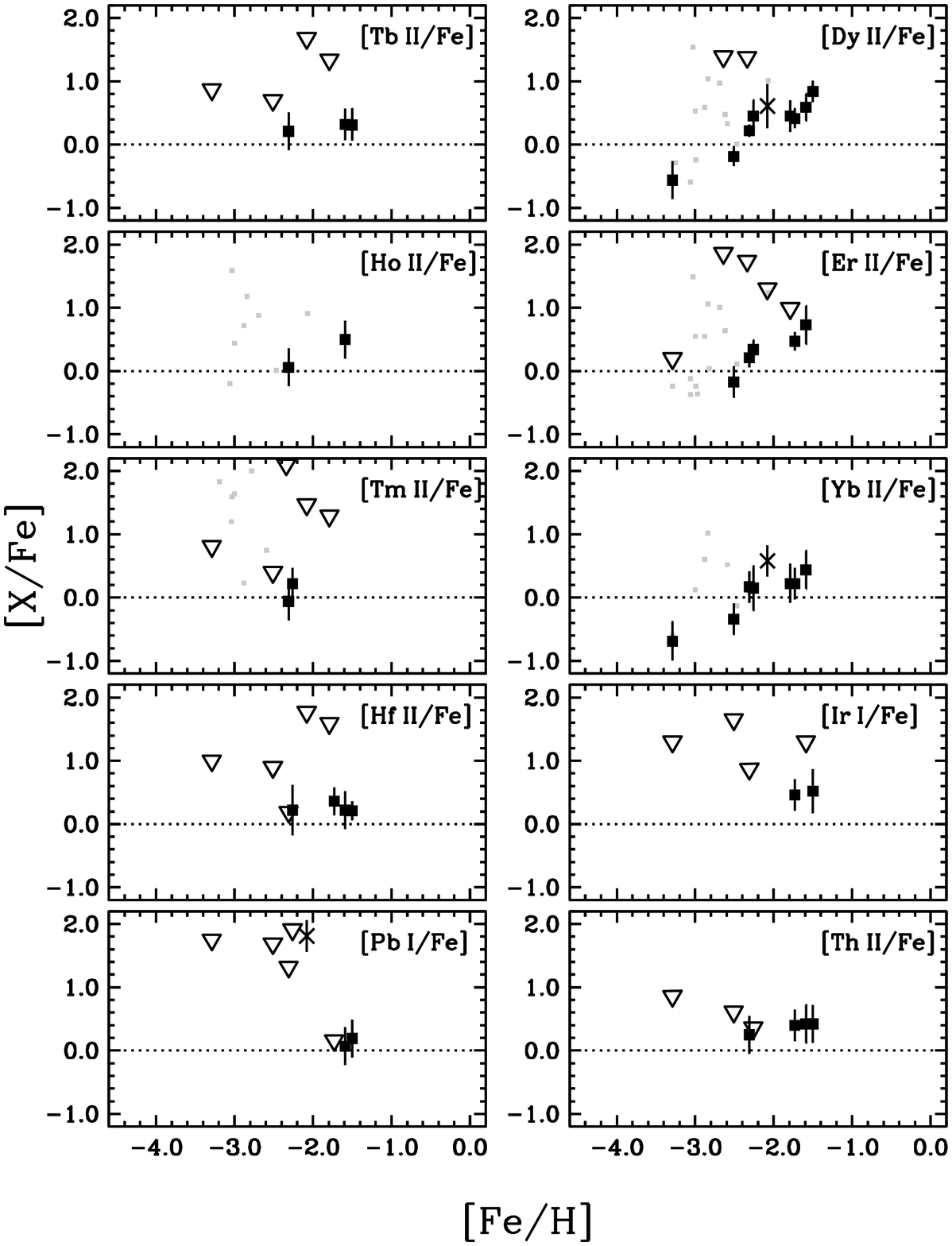}
\caption{
\label{abundplot5}
Abundance ratios [X/Fe] for Tb~\textsc{ii}--Th~\textsc{ii}
as a function of metallicity.
Symbols are the same as in Figure~\ref{abundplot1}.
}
\end{figure*}

Our derived [X/Fe] abundance ratios for the probable
stream members are shown in
Figures~\ref{abundplot1}--\ref{abundplot5}.
The stream members range in metallicity from 
$-3.4 \leq$~[Fe/H]~$\leq -1.5$, and they are not unique
with respect to the rest of the halo in this regard.
In these figures, \mbox{CS~29513--032} is indicated 
separately because its abundances for $Z \leq$~11 and
$Z \geq$~29 may not reflect their primordial (birth) abundances.
These abundances in \mbox{CS~29513--032}
are excluded from the star-to-star chemical
dispersions discussed in the following sections, but the
species with 12~$\leq Z \leq$~28 have been included.

\subsection{Carbon to Zinc}

The [C/Fe] ratios for the stream members are all sub-Solar by
a factor of 2--3.
[O/Fe] is super-Solar and similar to other metal-poor stars in the halo.
The [Na~\textsc{i}/Fe] and [Al~\textsc{i}/Fe] ratios are both sub-Solar,
suggesting that Na and Al have not been enriched by 
the CNO, NeNa, and MgAl cycles
like stars found in globular clusters.
The Al non LTE (NLTE) line formation 
corrections suggested by \citet{andrievsky08}
for the 3961\AA\ line would increase the [Al~\textsc{i}/Fe] ratios
to Solar or just slightly sub-Solar, but the overall
relative abundances are basically unchanged.
The star-to-star dispersions (standard deviation) are 0.29~dex for
[C/Fe], 0.07~dex for [Na~\textsc{i}/Fe], and 0.17~dex
for [Al~\textsc{i}/Fe].

The $\alpha$ elements (O, Mg, Si, Ca, and Ti) 
are all enhanced relative to Fe at levels similar to other stars 
in the halo.
The dispersion for [O~\textsc{i}/Fe] is 0.36~dex,
although this drops to 0.13~dex if only two stars (\mbox{BD$+$30~2611} 
and \mbox{HD~237846}) are excluded.
The dispersions are similarly small for 
[Mg~\textsc{i}/Fe] (0.10~dex) and 
[Ca~\textsc{i}/Fe] (0.07~dex);
the somewhat larger
dispersion for [Si~\textsc{i}/Fe] (0.16~dex)
and [Ti~\textsc{i}/Fe] (0.17~dex)
can be attributed to the $T_{\rm eff}$ correlations found in
Section~\ref{si}.
[Ti~\textsc{ii}/Fe] shows no such $T_{\rm eff}$ correlation,
so the larger scatter observed here (0.14~dex) may be genuine.
If we ignore [Si~\textsc{i}/Fe],
the stream members show no evolution in their [$\alpha$/Fe] ratios
over nearly 2~dex in [Fe/H].

The [K~\textsc{i}/Fe] ratios are super-Solar and show
no evolution over $-2.3 \leq$~[Fe/H]~$\leq -1.5$ within
the observational uncertainties.
\citet{ivanova00} and \citet{takeda09} computed 
NLTE corrections for the optical K~\textsc{i} resonance lines 
for metal-poor stars, and both groups found 
corrections of $-$0.20 to $-$0.35 over the evolutionary states
of our sample.
Thus our overall [K~\textsc{i}/Fe] ratio may need to be revised
downward by a factor of $\sim$~2 (as should the other metal-poor
stars illustrated in Figure~\ref{abundplot1}), 
but this should have minimal impact on the star-to-star dispersion
(0.21~dex).

The [Sc~\textsc{ii}/Fe] and [V~\textsc{ii}/Fe] ratios for the
stream members are roughly Solar.
Both exhibit moderate dispersions
(0.18 and 0.13~dex, respectively).
The large [V~\textsc{i}/Fe] dispersion (0.21~dex) is a consequence of the
$T_{\rm eff}$ dependence.

The [Cr~\textsc{i}/Fe] and [Cr~\textsc{ii}/Fe]
ratios show very small star-to-star dispersion
(0.05 and 0.07~dex, respectively),
and the and [Cr~\textsc{i}/Fe] ratios are $\approx$~0.3--0.4~dex
lower than the [Cr~\textsc{ii}/Fe] ratios.
\citet{sobeck07} reexamined the transition probabilities and
Solar abundance of Cr~\textsc{i}, finding that the 
[Cr~\textsc{i}/Fe] ratios were 0.15--0.20~dex lower than the
[Cr~\textsc{ii}/Fe] ratios.
Yet \citet{sobeck07} found no compelling evidence
that this discrepancy was due to NLTE effects, and 
their stellar sample suggested that the Cr~\textsc{i} and 
Cr~\textsc{ii} abundances may be more discrepant at lower metallicities;
our results qualitatively support this conclusion,
although this does not imply which set of stellar [Cr/Fe] ratios
(if either) should accurately reflect the true value.

The [Mn~\textsc{i}/Fe] ratios for the stream members are
sub-Solar by a factor of 2, show no evolution over the metallicity
range, and have a very small dispersion (0.11~dex).
This last attribute is not a consequence of our decision
to adjust the abundance of the resonance lines by $+$0.3~dex:
the dispersion in [Mn~\textsc{i}/Fe] derived from
only the higher-excitation lines is only 0.06~dex (8~stars).
Thus the small dispersion in [Mn~\textsc{i}/Fe] is intrinsic.

[Co~\textsc{i}/Fe], [Ni~\textsc{i}/Fe], and [Zn~\textsc{i}/Fe]
are all Solar (within 0.1~dex),
have small or moderate dispersions 
(0.07, 0.06, and 0.16~dex, respectively), and
show no evolution with metallicity (with the possible exception 
of Zn in \mbox{HD~237846}).
The [Cu~\textsc{i}/Fe] ratio was only derived for three stars
(plus \mbox{CS~29513--032}).
For these three stars, the [Cu~\textsc{i}/Fe] ratio is in 
good agreement with other stars in the halo, and the dispersion is
moderate, 0.20~dex.

\begin{figure}
\includegraphics[angle=270,width=3.4in]{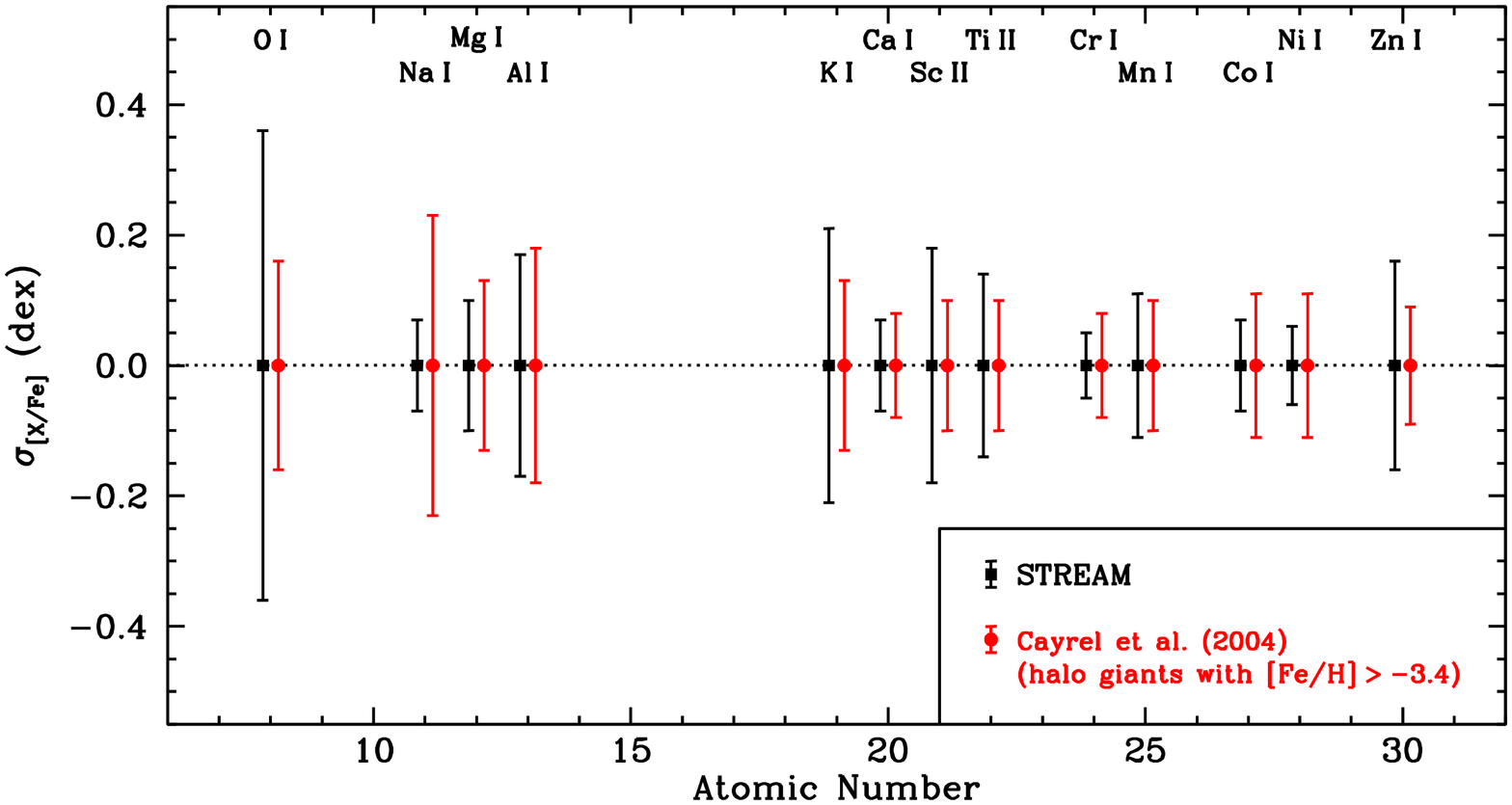}
\caption{
\label{dispersionplot}
The star-to-star dispersion in [X/Fe] ratios for the stream members
compared with the \citet{cayrel04} sample of metal-poor halo giants.
Only the halo giants with [Fe/H]~$> -3.4$ have been included.
Abundance ratios that show a dependence on $T_{\rm eff}$ in the 
stream members ([Si~\textsc{i}/Fe] and [Ti~\textsc{i}/Fe]) have
been excluded.
}
\end{figure}

How does the star-to-star dispersion observed in the stream members
compare with the rest of the halo?
In Figure~\ref{dispersionplot} we illustrate the dispersion in the
various [X/Fe] ratios and compare with the dispersion computed for
metal-poor red giants in the sample of \citet{cayrel04}.
Table~\ref{dispersiontab} lists these values.
The \citet{cayrel04} sample was optimized to measure 
the true cosmic dispersion 
(i.e., with observational uncertainties minimized)
of the very metal-poor end of the Galactic halo.
The stream spans a metallicity range of $-3.4 <$~[Fe/H]~$< -1.5$, so
we only compare with the metal-rich end of this sample
($-3.4 <$~[Fe/H]~$< -2.0$)
We do not attempt to compare abundance ratios that exhibit a
dependence on $T_{\rm eff}$ in our sample.
On average, the dispersion in the stream is comparable to 
or smaller than the dispersion of the two datasets of halo giants.
The implications of this point will be discussed further in 
Section~\ref{nature}.

\subsection{Strontium to Thorium}

In this section, we will ignore \mbox{CS~29513--032}, whose
heavy element abundances likely do not represent their 
initial values.
For the remaining stream stars,
most of the heavier elements follow similar patterns to the 
$\alpha$- and Fe-group elements, with one clear distinction:
the stream members with [Fe/H]~$> -2.2$ show 
a constant value of [X/Fe] (X stands for Ba and all heavier elements
in this case).
These values range from $\approx$~Solar (Ba, La) to 
$\approx$~3--4 times Solar (e.g., Eu, Gd, Dy, Er).  
The stream members with [Fe/H]~$< -2.2$ show gradually increasing
[X/Fe] ratios with increasing [Fe/H] 
that appear to culminate near the [X/Fe] ratios
of the more metal-rich stars.
The lighter elements Y and Zr show similar trends, but one star
(\mbox{CS~22876--040}, [Fe/H]~$= -2.34$) stands out with 
[Y~\textsc{ii}/Fe] and [Zr~\textsc{ii}/Fe] ratios 
$\approx$~0.6~dex higher than the other stars with [Fe/H]~$= -2.3$.
Otherwise the [Y~\textsc{ii}/Fe] and [Zr~\textsc{ii}/Fe] ratios
for the stars with [Fe/H]~$< -2.2$ show a gradual increase
of [X/Fe] with increasing [Fe/H].
[Y~\textsc{ii}/Fe] and [Zr~\textsc{ii}/Fe] show dispersions
of 0.10 and 0.15~dex among the stars with [Fe/H]~$> -2.2$.
The [Sr~\textsc{ii}/Fe] ratio shows a moderate
dispersion (0.31~dex) at all metallicities.
The [Mo~\textsc{i}/Fe] ratio shows no evolution with metallicity
and has a dispersion of 0.09~dex.

Production of the elements heavier than the Fe-group occurs primarily
by successive neutron ($n$) captures on existing nuclei.
The resulting abundance patterns are largely determined by the
rate of neutron captures, either slow ($s$) or rapid ($r$)
relative to the nuclear $\beta^{-}$ decay rates.
The general abundance patterns of these two processes are 
relatively well known, and they can be readily identified 
when one process dominates the production of the heavy isotopes.
Relatively large amounts of material tend to build up when 
either \ncap\ process encounters closed nuclear shells
at $N$ (or $Z$)~$=$~50, 82, or 126;
these relative overabundances are commonly referred to as the
1$^{\rm st}$, 2$^{\rm nd}$, and 3$^{\rm rd}$ peaks, respectively.

One of the surprising results of many detailed investigations
of \ncap\ abundances in metal-poor stars over the last 15 years has 
been the near-perfect match between the 
stellar distribution for the rare earth elements (La--Yb),
3$^{\rm rd}$ peak elements (Os--Pt), and actinides (Th)
and the scaled-Solar \rpro\ distribution.
This agreement has only improved with better atomic data
\citep{sneden09}.
This pattern is observed in many stars in different
populations that must be enriched by separate events.
The constant $n$-capture-element to $n$-capture-element ratios
do not extend to material at the 1$^{\rm st}$ \rpro\ peak
(e.g., \citealt{truran02}).

\begin{figure}
\includegraphics[angle=0,width=3.4in]{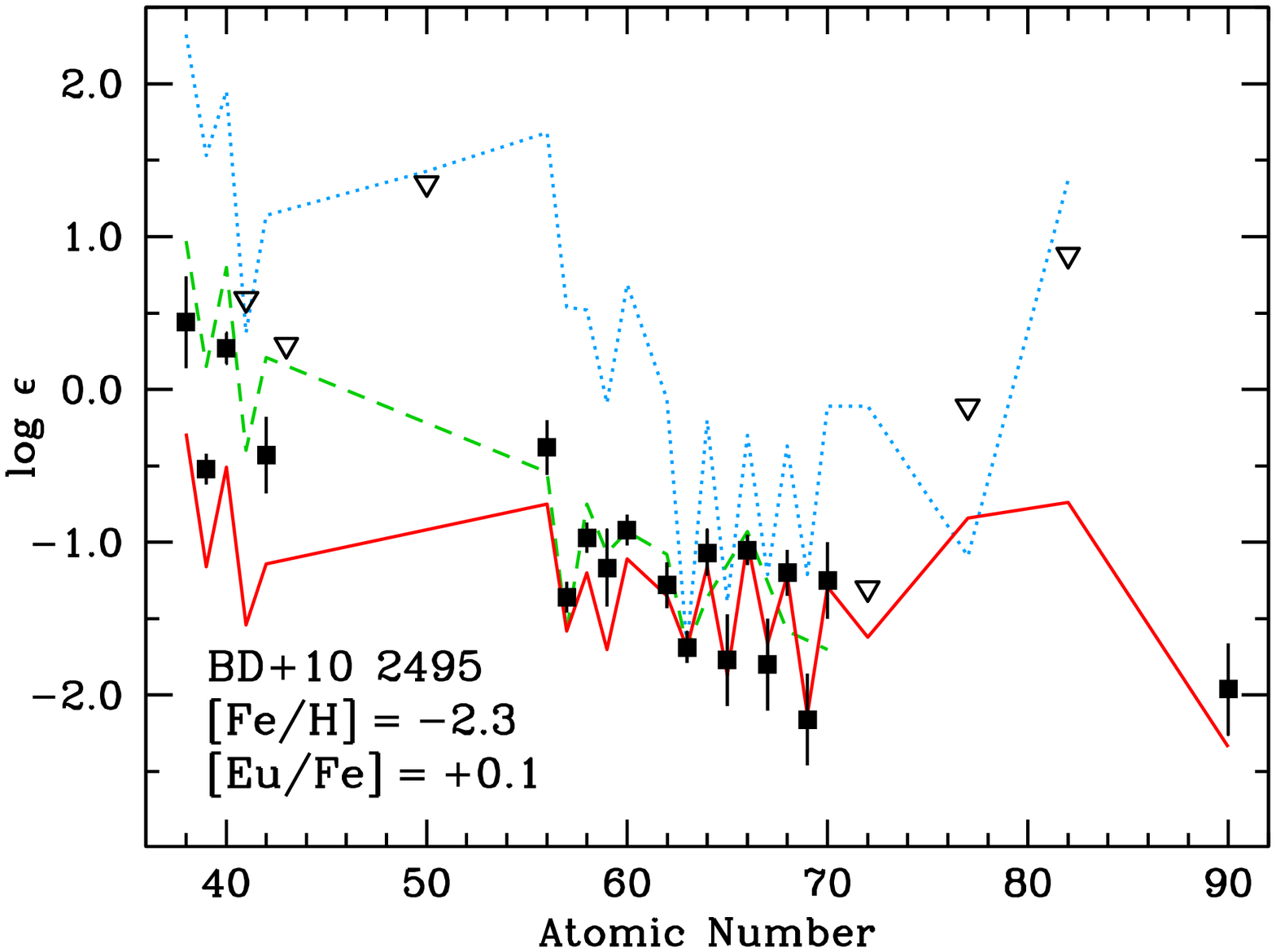}
\caption{
\label{bdp102495plot}
Neutron-capture abundances for \mbox{BD$+$10~2495}.
Squares indicate measurements and open triangles indicate upper limits.
Abundances for \mbox{CS~22892--052} (the template for the
main component of the \rpro, shown as a solid line) have been adopted from
\citet{sneden03a,sneden09}, abundances for \mbox{HD~122563}
(the template for the weak component of the \rpro, dashed line)
have been adopted from \citet{honda07},
and the \spro\ abundances (dotted line) are taken from the
Solar metallicity stellar model of \citet{arlandini99}.
The curves are normalized to the Eu abundance.
}
\end{figure}

In Figure~\ref{bdp102495plot} we show the abundance distribution
for the \ncap\ material in one stream star, \mbox{BD$+$10~2495}.
Three \ncap\ enrichment templates are shown for comparison:
the main component of the \rpro\ (exemplified by the 
well-studied star \mbox{CS~22892--052}),
the main component of the \spro\ (exemplified by the
Solar-metallicity model of the \spro\ from \citealt{arlandini99}),
and the so-called weak component of the \rpro\
(exemplified by the well-studied star \mbox{HD~122563}).
The curves are normalized to one another at the Eu abundance.
For the heaviest elements ($Z \geq$~62), the abundance pattern
very clearly follows the main component of the \rpro---\textit{even 
though the [Eu/Fe] ratio is only $+$0.1 relative to the Solar ratio}.
For the light rare earth elements (56~$\leq Z \leq$~60),
the abundances lie near but slightly above 
the main component of the \rpro\ and
close to the weak component of the \rpro.
The light \ncap\ elements (38~$\leq Z \leq$~42)
also fall between the weak and main components of the \rpro,
though usually tending towards the weak component.
A pure \spro\ is clearly ruled out.

\begin{figure*}
\includegraphics[angle=0,width=7.0in]{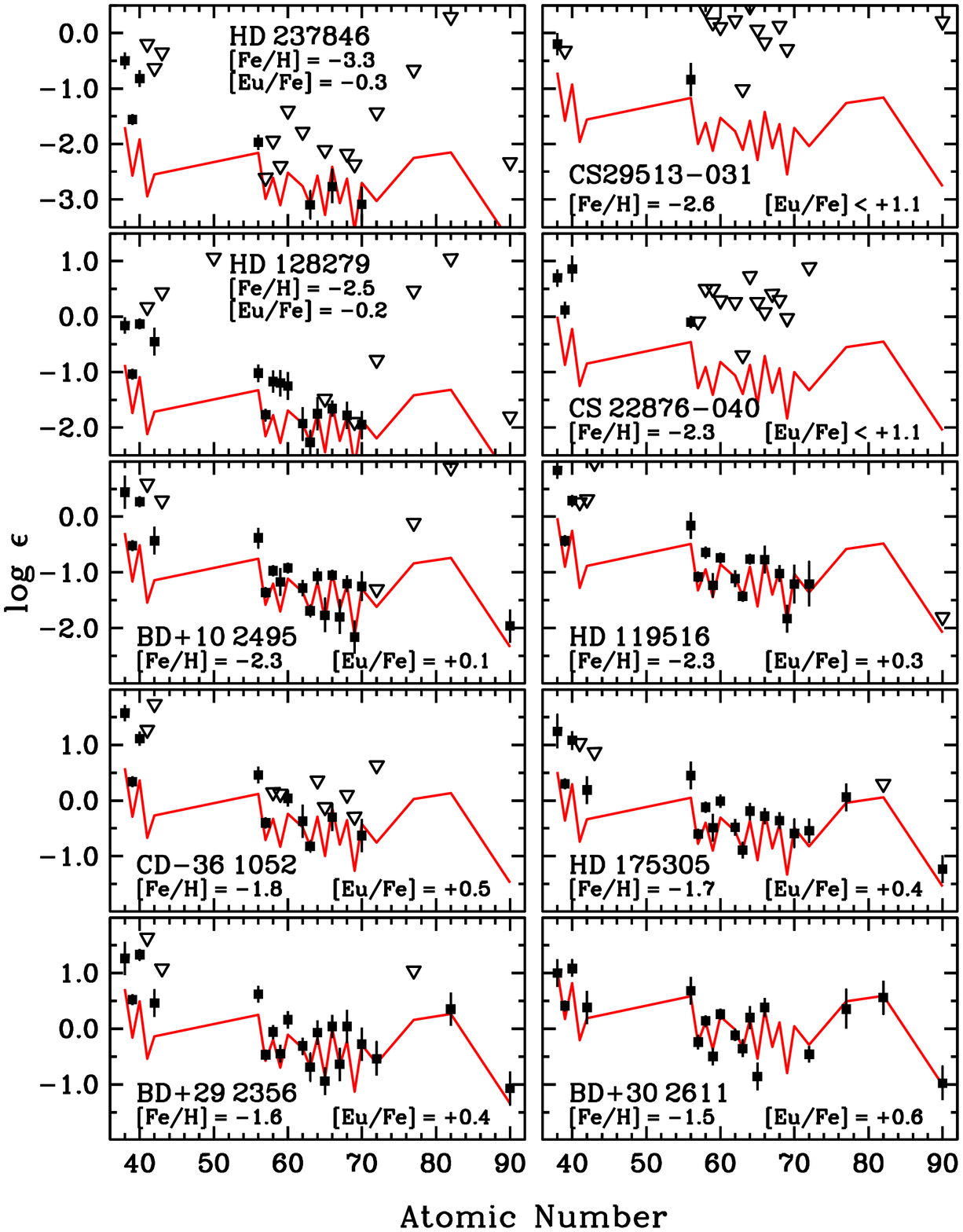}
\caption{
\label{ncapplot}
Neutron-capture abundances for ten of the stream members.
(Regarding the other two members: 
\mbox{CS~29513--032} is analyzed in detail in the Appendix, and
only the Sr~\textsc{ii} and Ba~\textsc{ii} abundances
were derived for \mbox{CS~22948--093}.)
Squares indicate measurements and open triangles indicate upper limits.
Abundances for \mbox{CS~22892--052} (the template for the
main component of the \rpro, shown as a solid line) have been adopted from
\citet{sneden03a,sneden09}.
The curve is normalized to the Eu abundance in each star,
with the exception of \mbox{CS~29513--031} and \mbox{CS~22876--040}
(Eu upper limit),
where the curve
scaling is approximately matched to the Ba abundance offset
observed in the remaining stars.
}
\end{figure*}

Similar plots are shown for ten stream members in Figure~\ref{ncapplot},
except here only the template for the main component of the
\rpro\ is shown to reduce clutter.
Clearly, in all cases where at least a few \ncap\ elements can be 
detected, the enrichment pattern is virtually identical to
that in \mbox{BD$+$10~2495}.

Stars enriched in \spro\ material by companions that passed through the
AGB phase of evolution are also enriched in C \citep{sneden03b}, 
which is not the case for any of the stream members.
Furthermore, no stream members have high [Pb/Fe] ratios.
An enhanced Pb ($Z =$~82) abundance will be the first 
detectable signature of the \spro\ at low metallicity.
At low metallicity, due to the higher ratio of neutrons 
to Fe-group seed nuclei for the \spro, 
models predict that a low metallicity \spro\ produces large
Pb/Fe and Pb/2$^{\rm nd}$ peak ratios 
relative to higher metallicity \spro\ models
(e.g., \citealt{gallino98}).
This has been confirmed by a number of observational studies
(e.g., \citealt{vaneck03}, \citealt{ivans05}).
If the observed \ncap\ abundance pattern in the stream stars
were to result from the combination of material produced in the
main component of the \rpro\ and the main component of the \spro,
we would expect to detect large Pb/2$^{\rm nd}$ peak ratios
in these stars.
In the two stars where Pb is detected (\mbox{BD$+$29~2356} and
\mbox{BD$+$30~2611})---the two most metal-rich stars 
in our stream sample---the Pb abundance clearly matches 
the pure \rpro\ pattern.\footnote{
Pb has not been detected in \mbox{CS~22892--052}. 
The Pb abundance indicated by the \mbox{CS~22892--052} curve
in Figure~\ref{ncapplot} has been derived from the measurements
of \citet{roederer09b}, who detected Pb in a number of other
stars with $-2.2 <$~[Fe/H]~$< -1.4$.
No hint of \spro\ contamination could be detected in this sample.
}
The Pb upper limit in \mbox{HD~175305} also strongly suggests
an \rpro\ origin.

We conclude that the \ncap\ material was produced by the main 
component of the \rpro\ 
with significant contributions from the weak component of the \rpro\
at the 1$^{\rm st}$ peak and light end of the rare earth domain.
There is no evidence that any of the material in the
stream stars originated in the \spro.

\section{Discussion}
\label{discussion}

\begin{figure*}
\includegraphics[angle=0,width=7.0in]{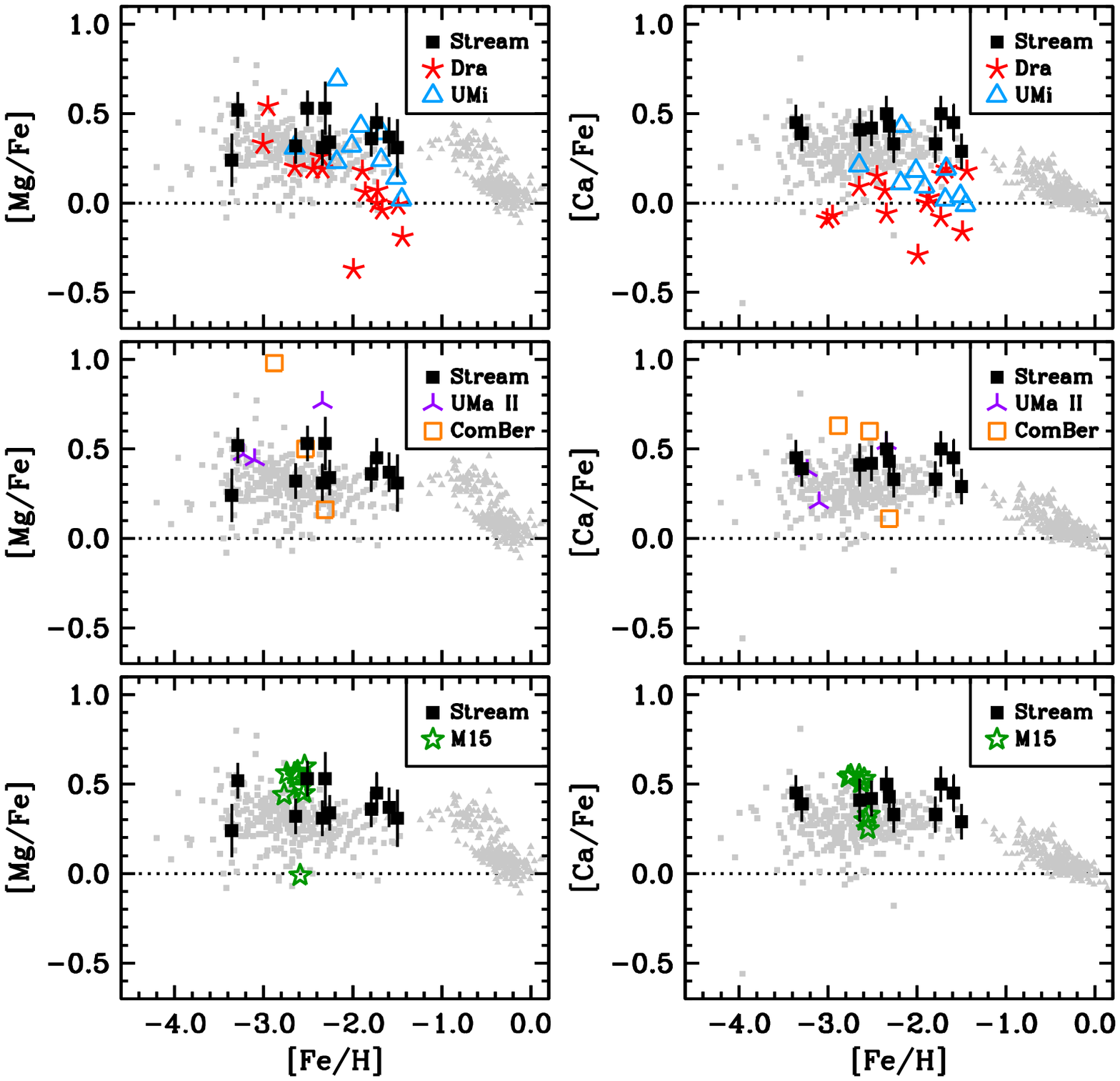}
\caption{
\label{detailplot1}
Abundance ratios [Mg/Fe] and [Ca/Fe] for the stream stars and
select other stellar populations.
Gray triangles indicate thin and thick disc stars \citep{reddy03,reddy06}.
Gray squares indicate field halo stars
\citep{cayrel04,barklem05,cohen08,lai08}.
Abundances for the Ursa Minor dSph are taken from
\citet{shetrone01}, \citet{sadakane04}, and \citet{aoki07}.
Abundances for the Draco dSph are taken from
\citet{shetrone01}, \citet{fulbright04}, and \citet{cohen09}.
Abundances for the Coma Berenices and Ursa Major II uFds are
taken from \citet{frebel09}.
Abundances for globular cluster M15 are taken from \citet{sobeck10}.
The Solar ratio is indicated in each panel by the dotted line.
}
\end{figure*}

\subsection{The Chemical Nature of the Stream}
\label{nature}

\begin{figure*}
\includegraphics[angle=0,width=7.0in]{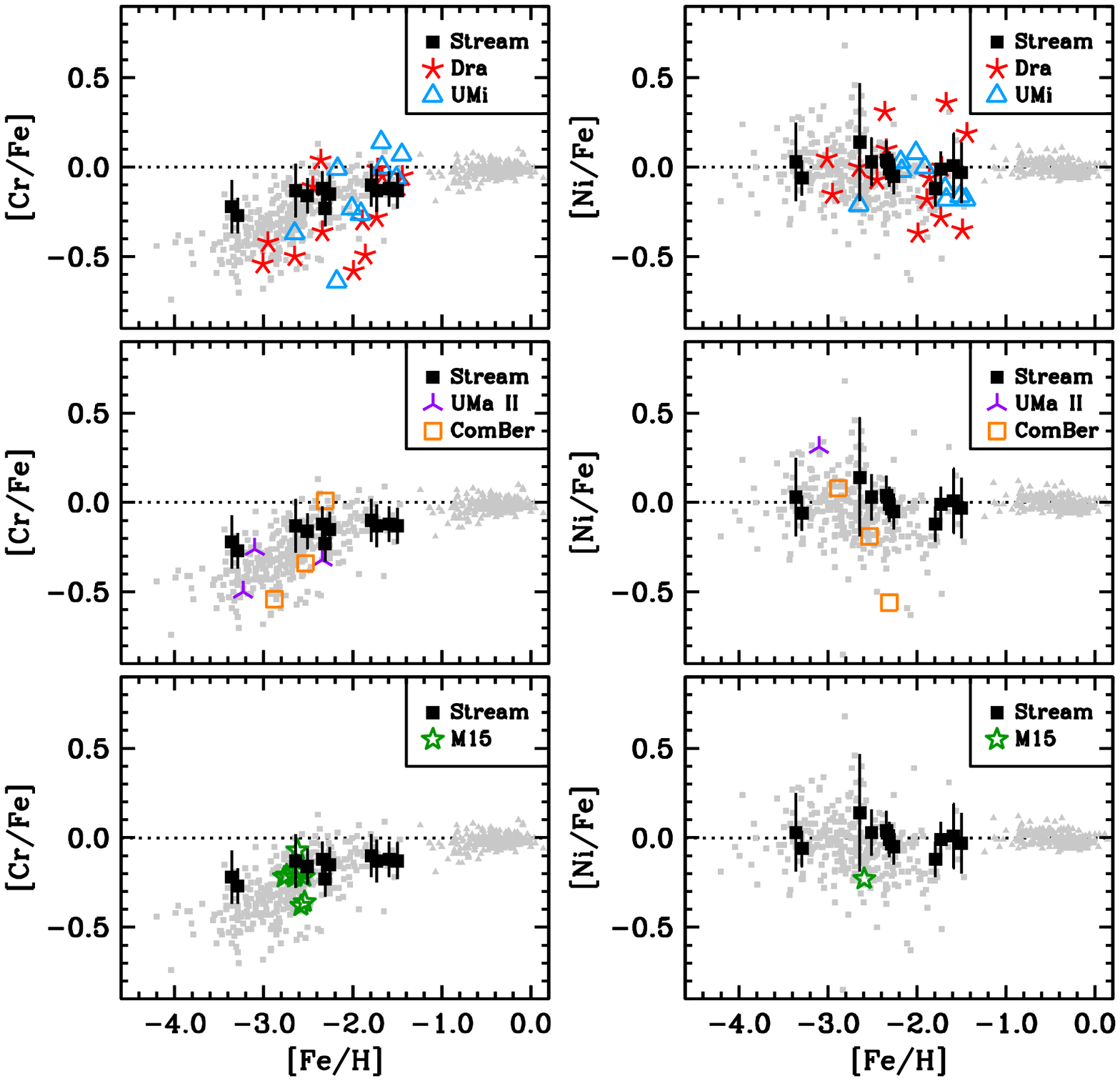}
\caption{
\label{detailplot2}
Abundance ratios [Cr/Fe] and [Ni/Fe] for the stream stars and
select other stellar populations.
Symbols are the same as in Figure~\ref{detailplot1}.
}
\end{figure*}

The abundance pattern in the stream stars appears to be the 
result of massive Type~II supernovae,
characterized by sub-Solar [C/Fe] and enhanced [$\alpha$/Fe] ratios.
The [Fe/H] abundances span almost 2~dex,
but the [X/Fe] ratios for the $\alpha$ and Fe-group elements
are basically unchanged.
The \ncap\ elements are the lone exception, showing Solar or sub-Solar
ratios at the lowest metallicities in the stream but increasing to
a super-Solar plateau at the highest metallicities.
One proposed site for the \rpro\ is the 
high-entropy wind of $\sim$~8--10~$M_{\odot}$ Type~II supernovae,
which may be capable of producing both the weak and main components
of the \rpro\ (e.g., \citealt{farouqi09}).
This scenario seems to imply that the lowest metallicity stream stars
([Fe/H]~$\lesssim -2.2$)
were formed from gas
polluted by more massive Type~II supernovae,
while the more metal-rich stars ([Fe/H]~$\gtrsim -2.2$)
were formed from gas polluted by 
less-massive Type~II supernovae.
Subsequent generations of supernovae then continue to enrich the
ISM, 
but star formation appears to have been truncated before the
yields of Type~Ia supernovae or AGB stars contributed significantly to the
chemical inventory.
This would also imply that whatever site is responsible
for producing the \rpro\ does not produce a 
difference in the light element ($Z \leq$~30) production ratios
from the sites that do not produce an \rpro.

\begin{figure*}
\includegraphics[angle=0,width=7.0in]{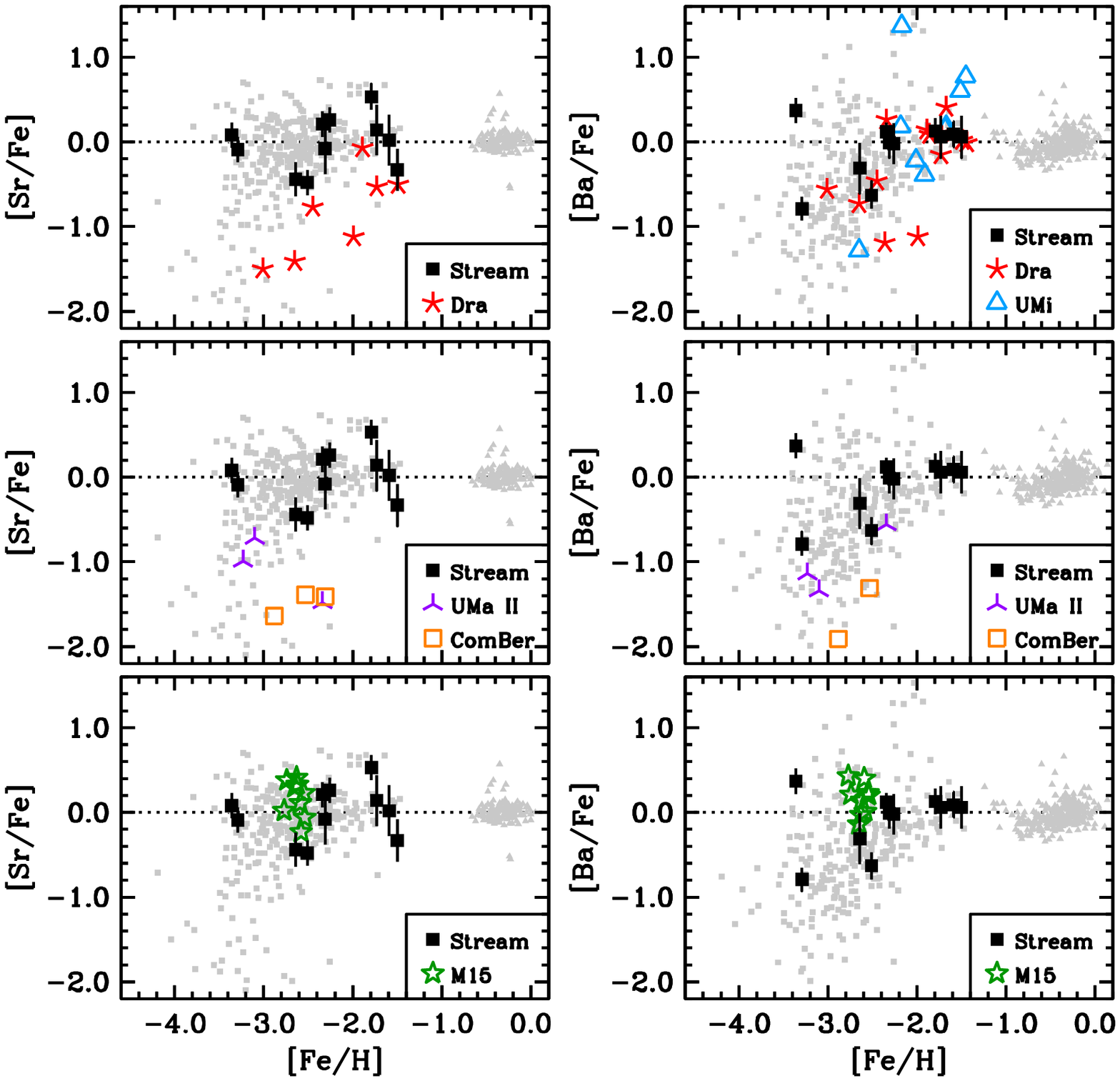}
\caption{
\label{detailplot3}
Abundance ratios [Sr/Fe] and [Ba/Fe] for the stream stars and
select other stellar populations.
Symbols are the same as in Figure~\ref{detailplot1}.
}
\end{figure*}

In Figures~\ref{detailplot1}, \ref{detailplot2}, and \ref{detailplot3}
several abundance ratios for the stream stars
are compared with analogous abundance ratios of
other stellar populations.
These include field stars in the thin and
thick disc populations, halo stars, two dSph galaxies
(Draco and Ursa Minor),
two uFds (Coma Berenices and Ursa Major~II),
and the metal-poor globular cluster M15.
At a given metallicity, 
the stream shows an equal or smaller star-to-star dispersion in [X/Fe] 
when compared with stars in the dwarf galaxies.
Furthermore, the dwarf galaxies show evolution in their
[X/Fe] ratios as a function of metallicity
(e.g., Mg or Cr);
the stream only shows such evolution for the \ncap\ elements.
The \ncap\ elements show significantly larger dispersion in these
dwarf galaxies than in the stream, with the dwarf galaxy
dispersions often reaching 1--2~dex at a single metallicity.

The Sgr dSph (not illustrated in these Figures) 
is much more metal-rich than the stream and shows clear 
evidence for a large dispersion and evolution 
of several [X/Fe] ratios with metallicity (e.g., 
\citealt{monaco05,monaco07,sbordone07}).
\citet{majewski03} argue that this particular stream cannot be related
to Sgr debris because its $R_{\rm apo}$ is too small and its
$L_{z}$ angular momentum is too high. 
Furthermore, Sgr contributes less than 1\% of the 
evolved halo stars in the Solar neighborhood, and the Sgr debris
would not have significantly impacted the kinematic studies
that have detected the presence of this stream.
We reaffirm this conclusion 
on the basis of the composition of the stream stars.

The stream does not resemble a globular cluster in that it shows
a range of metallicities spanning nearly 2~dex, whereas
globular clusters, such as M15, show minimal or no
internal metallicity spread
(except for $\omega$ Centauri).
The stream also does not resemble a dSph or uFd system,
whose [X/Fe] ratios show much larger star-to-star dispersion
and evolution of the [X/Fe] ratios as a function of [Fe/H]
(insofar as the true chemical dispersion can be estimated 
for the uFd systems from measurements of only 3~stars in 
each of two systems; \citealt{frebel09}).
The luminous dSph systems spend---at most---a very
small fraction of their lives in the Solar neighborhood
\citep{roederer09a},
so these particular systems would not be expected to 
have spawned the stream.
The stream has the same chemical dispersion
as the rest of the local halo, and it is
reasonable to hypothesize that, in principle, 
a significant fraction of field halo stars
(at least those currently in the Solar neighborhood)
could form in progenitor systems like 
that from which the stream originated.

Two stars in the stream have [Fe/H]~$< -3.0$, \mbox{HD~237846}
([Fe/H]~$= -3.3$) and \mbox{CS~22948--093} ([Fe/H]~$= -3.4$).
While it is possible that these stars have no association
with the stream yet by pure chance have similar kinematics,
there is no a priori reason
to exclude them from membership on these grounds.
Multiple enrichment events spanning multiple stellar generations 
are required to enrich an unpolluted 
ISM to a metallicity of [Fe/H]~$= -1.5$, the metal-rich end of our
stream stars.
Therefore, it is reasonable to expect that at least a few stars
with [Fe/H]~$< -3.0$ formed in the stream's progenitor system
throughout the enrichment process.
Furthermore, the [X/Fe] abundance ratios in these two stars are
generally consistent with the other stream members.
In the last few years a number of stars with [Fe/H]~$< -3.0$ have
been identified in dwarf galaxies
\citep{cohen09,frebel09,geha09,kirby09,norris09}.
If a significant fraction of the stellar halo is postulated 
to be formed by the accretion of satellites 
like the stream's (unidentified) progenitor,
we should not be surprised to encounter stream stars with [Fe/H]~$< -3.0$.

\subsection{Stellar Ages from Nuclear Cosmochronometry}
\label{ages}

We have detected Th in four stream members, which permits
us to estimate the ages of these stars by comparing the abundance of Th
(which is radioactive with a halflife
$t_{\rm 1/2} [^{232}{\rm Th}] = 14.05\,\pm\,$0.06~Gyr;
\citealt{audi03})
to another stable element produced in the same nucleosynthesis event.
Th can only be produced by the main \rpro,
and in this stream the Eu has also only produced by the main \rpro.
By referencing the current Th/Eu ratio against the expected production ratio
(e.g., \citealt{kratz07}) an age for the \rpro\ material (and an upper limit
for the age of the stars themselves) can be calculated.
The mean age for the four stars in the stream is 3.1~$\pm$~7.9~Gyr, and
this uncertainty is dominated by the star-to-star dispersion 
in Th/Eu.\footnote{
In this case the uncertainty only marginally improves when considering 
ages for a single star.
The long halflife of $^{232}$Th limits the age resolution to only 1~Gyr
per 0.021~dex of uncertainty in Th/Eu.
For an uncertainty of 0.20~dex, then, the uncertainty in the age
rises to 9.5~Gyr.
Even when the observational scatter has been minimized, 
the uncertainties from the production ratios still only allow a precision
of $\sim$~2--5~Gyr in the absolute age \citep{frebel07}.
Relative age determinations avoid the uncertainties in the production ratios,
which must be calculated from theory.
Selecting a reference element other than Eu (or a mean of several) will 
not affect the result significantly, provided that some of the rare earth
elements are treated with due caution.
See further discussion in Section~6 of \citet{roederer09b}.
}

\begin{figure}
\includegraphics[angle=0,width=3.4in]{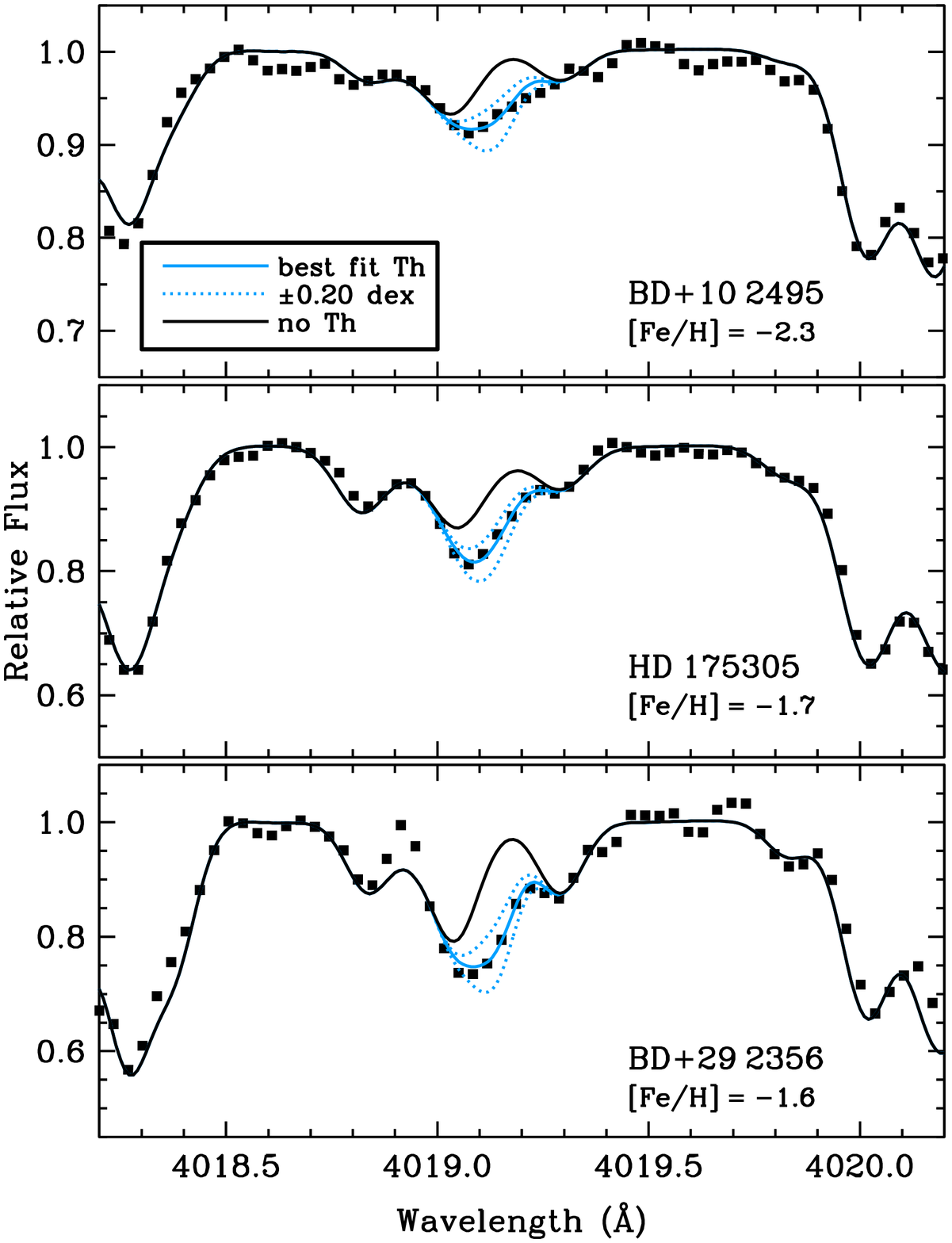}
\caption{
\label{thoriumplot}
Syntheses of the Th~\textsc{ii} line in the three stars
where we have used this line to derive the Th abundance.
}
\end{figure}

This large uncertainty prevents us from making any meaningful
statements about the age of the stream, and the mean age is
uncomfortably small for metal-poor stars.
The Th~\textsc{ii} abundance has been derived from a single transition
in each of the stream stars, and this line is always weak ($\sim$~5~m\AA)
and blended with other features.  
The Th/Eu ratios for three of the stars (\mbox{BD$+$10~2495}, 
\mbox{BD$+$29~2356}, \mbox{HD~175305}) are higher than the fourth
(\mbox{BD$+$30~2611}); the Th abundance in the former three stars 
was derived from the 4019\AA\ line, while the 4094\AA\ line was used
in the latter due to severe blending at the 4019\AA\ line.
\citet{roederer09b} found no systematic difference in the Th abundance derived
from these lines.
Our syntheses of the 4019\AA\ Th~\textsc{ii} line are shown in 
Figure~\ref{thoriumplot}.

There are currently four metal-poor field stars with exceptionally high 
Th/Eu ratios.
In the one star in this class with detected U, \mbox{CS~31082--001}, the 
U/Th ratio gives a reliable age, while the U/$r$ and Th/$r$ ratios
(where $r$ is any reference element)
predict negative ages \citep{hill02,plez04}.
This so-called ``actinide boost'' \citep{schatz02}, which seems only
to affect the \rpro\ material heavier than the 3$^{\rm rd}$ \rpro\ peak
\citep{roederer09b}, is characterized by low Pb and enhanced Th and U 
(relative to the majority of \rpro\ enriched metal-poor stars).
This nucleosynthetic idiosyncrasy is unexplained by current models of
the \rpro.
We do not detect U in any of these stars, nor can we place a 
meaningful upper limit on its abundance.
The Pb abundance in two of these stars is normal (i.e., consistent with 
no actinide boost), and the Pb upper limit in the other
two cannot exclude a normal Pb abundance.
The actinide boost phenomenon does not seem to be able to 
explain the high Th/Eu ratios in these stars.

It is possible that these stars (and the \rpro\ material in them) may
be younger than the rest of the halo,\footnote{
Curiously, isochrones computed for ages 6--9~Gyr are a better fit
to the turnoff and subgiant stream stars in Figure~\ref{cmd}
than isochrones computed for 10--13~Gyr.}
but we urge caution in
interpreting the ages calculated from abundances derived from
a single, weak, and blended Th feature in each of these stars.
Postulating that some fraction of the Eu was formed via the weak
\rpro\ (instead of the main component of the \rpro, which produces
the Th) only exacerbates the age discrepancy.

\section{Conclusions}
\label{conclusions}

We have performed a detailed abundance analysis of 12 metal-poor 
halo field stars whose kinematics suggest they are
members of the stellar stream first discovered by \citet{helmi99b}.
These stars exhibit a range of metallicity
($-3.4 \leq$~[Fe/H]~$\leq -1.5$)
but are otherwise chemically homogeneous for elements with $Z \leq 30$.
The [$\alpha$/Fe] ratios are enhanced to levels typical for 
stars in the local Galactic halo
(e.g., [Mg/Fe] or [Ca/Fe]~$= +0.4$).
The star-to-star dispersion in [X/Fe] is very small and the
same as found for other halo field stars
(e.g., the sample of metal-poor giants analyzed by \citealt{cayrel04}).
The \ncap\ elements are deficient 
at the lowest metallicities
(e.g., [Ba/Fe]~$= -0.8$ or [Eu/Fe]~$= -0.3$) 
but increase and plateau at the highest metallicities
(e.g., [Ba/Fe]~$= 0.0$ or [Eu/Fe]~$= +0.4$).
The \ncap\ elements are clearly produced by the 
main and weak components of the \rpro, and there is no evidence
for enrichment by the \spro.
These enrichment patterns can be produced by Type~II core-collapse
supernovae, implying that star formation in the stream progenitor
was truncated before the products of Type~Ia supernovae or AGB 
stars enriched the ISM.

\begin{figure}
\includegraphics[angle=0,width=3.4in]{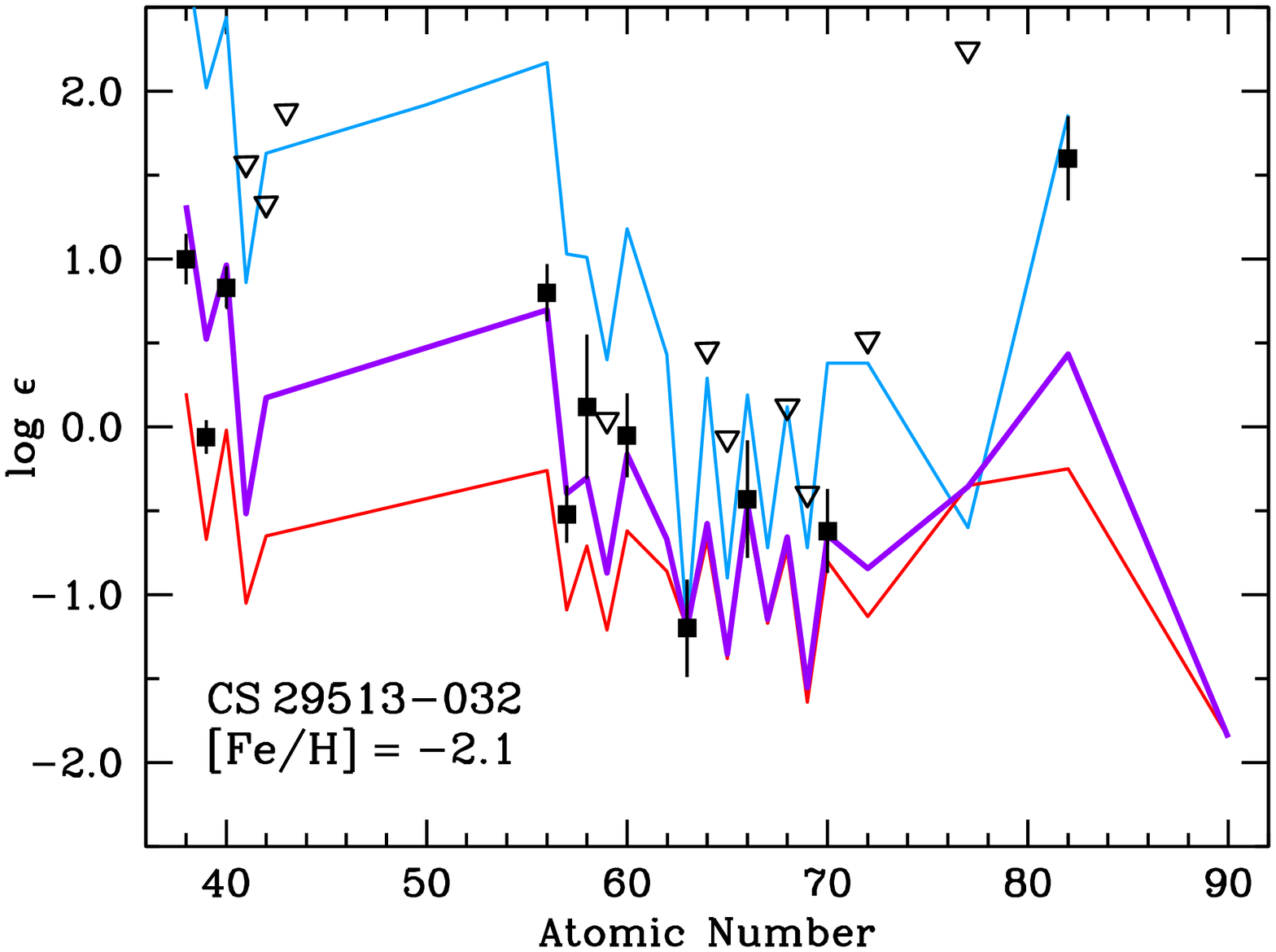}
\caption{ 
\label{cs29513m032}
Derived neutron-capture abundances in \mbox{CS~29513--032}.
Symbols are the same as in Figure~\ref{ncapplot}.
The red curve indicates the \rpro\ standard star \mbox{CS~22892--052},
the blue curve indicates the Solar metallicity \spro\ predictions
of \citet{arlandini99}, and
the bold purple curve indicates a linear combination of the two
other curves that approximately matches the derived 
1$^{\rm st}$ peak and rare earth abundances. 
All curves are normalized to the derived Eu abundance.
}
\end{figure}

We find two extremely metal-poor stars ([Fe/H]~$= -3.3$ and $-3.4$)
in the stream, suggesting that
whatever progenitor produced the stream was capable of producing
extremely metal-poor stars like those observed in the stellar halo
and a handful of dwarf galaxies.
The stream stars span a range of metallicities,
unlike individual Galactic globular clusters, 
which have no significant internal metallicity spread.
The stream also exhibits smaller star-to-star chemical dispersion 
than the Milky Way dwarf galaxies.
We cannot identify a direct progenitor of the stream, but
our results support the notion that a significant fraction of the
Milky Way stellar halo can form from accreted systems.

\acknowledgments

I.U.R.\ wishes to thank J.\ Cowan and D.\ Yong for enlightening
discussions, A.\ McWilliam for providing the model atmosphere
interpolation code, and J.\ Sobeck for sharing abundance results
in advance of publication.
We thank the referee for several helpful suggestions 
that have improved this manuscript.
This research has made use of the 
NASA Astrophysics Data System (ADS),
NIST Atomic Spectra Database, 
SIMBAD database (operated at CDS, Strasbourg, France),
and the Milky Way Spheroid Substructure database\footnote{
http://www.rpi.edu/\~{}newbeh/mwstructure/MilkyWaySpheroidSubstructure.html}.
Funding for this project has been generously provided by 
the U.~S.\ National Science Foundation
(grants AST~06-07708 and AST~09-08978 to C.S.).

%
{\it Facilities:} 
\facility{Magellan:Clay (MIKE)}
\facility{Smith (2dCoude)}

\appendix

\section{CS~29513--032}
\label{cs29513m032text}

\mbox{CS~29513--032} shows clear enrichment by the \spro, with
large overabundances of [C/Fe]~$= +0.6$ and [Pb/Fe]~$= +1.8$.
The Ba is also enhanced, with [Ba/Fe]~$= +0.8$ and [Ba/Eu]~$= +0.4$.
In contrast, the other stream members---enriched by the \rpro---have 
$\langle$[Ba/Eu]$\rangle = -0.35$ ($\sigma = 0.13$).
Low-metallicity stars on the AGB (or stars polluted by their 
nucleosynthetic products)
are also predicted to be Na-enhanced, 
which has been observationally confirmed 
(e.g., \citealt{ivans05,roederer08}).
\mbox{CS~29513--032} is also Na-enhanced, with
[Na/Fe]~$\approx +0.15$ 
(approximately corrected for NLTE; e.g., \citealt{andrievsky07}).
The \ncap\ enrichment pattern of \mbox{CS~29513--032} is illustrated 
in Figure~\ref{cs29513m032}.
In contrast to the weak and main \rpro\ enrichment seen in 
Figure~\ref{bdp102495plot}, the Pb abundance lies far above
the Pb abundance predicted from the Pb/Eu \rpro\ ratio.

We can approximately fit the observed abundance pattern
in \mbox{CS~29513--032} by taking linear combinations of the
$s$- and \rpro\ template abundance patterns, as indicated by the bold
curve in Figure~\ref{cs29513m032}.
This predicted distribution (normalized to the Eu abundance)
provides a reasonable fit to the
1$^{\rm st}$ peak and rare earth elements in 
\mbox{CS~29513--032}.
The derived Pb abundance is still higher than the predicted Pb
abundance because the prediction is based on a Solar-metallicity
AGB model.
According to the observed and predicted [Pb/Ba] ratios for a
1.5~$M_{\odot}$ AGB model\footnote{
This is representative of the typical masses of stars on the AGB
that are likely sites for the \spro\ 
(e.g., \citealt{bisterzo09}).
} 
presented in Figure~20 of \citet{sneden08}, one might expect
a [Pb/Ba] ratio of $\sim +1.0$ ($\pm \sim 0.5$) at [Fe/H]~$= -2.0$.
This translates to [Pb$_{\rm [Fe/H]=-2}$/Pb$_{\rm [Fe/H]=0}$]~$\sim +1.0$.
Indeed, the derived Pb abundance in \mbox{CS~29513--032} is
roughly 1~dex higher than the predicted Pb abundance based on the
Solar \spro\ model.
Even in the absence of detailed calculations, the high [Pb/Fe] ratio
in \mbox{CS~29513--032} clearly indicates an \spro\ origin.

\mbox{CS~29513--032} is a subgiant and has not passed 
through the AGB phase of evolution
where the \spro\ is expected to occur.
We speculate that
the \spro\ material observed in this star was produced in another
low-metallicity star in the AGB phase,
likely a binary companion that has long since faded from view.

\clearpage
\input{tab1}

\input{tab2}

\clearpage
\input{tab3}

\clearpage
\input{tab4-stub}

\input{tab5}

\input{tab6}

\clearpage
\input{tab7}

\clearpage
\input{tab8}

\clearpage
\input{tab9}

\clearpage
\input{tab10}

\clearpage
\input{tab11}

\end{document}

%% file: tab1.tex
\begin{deluxetable}{lcccccc}
\tablecaption{Log of Observations
\label{obstab}}
\tablewidth{0pt}
\tabletypesize{\scriptsize}
\tablehead{
\colhead{Star} &
\colhead{Telescope/} &
\colhead{Exp.} & 
\colhead{Date} &
\colhead{UT mid-} &
\colhead{Heliocentric} &
\colhead{Heliocentric} \\
\colhead{} &
\colhead{Instrument} &
\colhead{time (s)} &
\colhead{} &
\colhead{exposure} &
\colhead{JD} &
\colhead{RV (\kmsec)} }
\startdata
BD$+$10~2495	& McDonald-Smith/2dCoude  & 5600   & 2009 May 06 & 06:23 & 2454957.771	 & $+$250.8	\\
BD$+$10~2495	& McDonald-Smith/2dCoude  & 3600   & 2009 Jun 10 & 04:31 & 2454992.690   & $+$250.8     \\
BD$+$29~2356    & McDonald-Smith/2dCoude  & 5400   & 2009 Jun 11 & 05:45 & 2454993.741   & $-$207.3     \\
BD$+$30~2611	& McDonald-Smith/2dCoude  & 2700   & 2009 May 06 & 07:43 & 2454957.826	 & $-$280.4	\\
BD$+$30~2611	& McDonald-Smith/2dCoude  & 2300   & 2009 Jun 10 & 05:25 & 2454992.729   & $-$280.0     \\
BD$+$30~2611	& McDonald-Smith/2dCoude  & 2500   & 2009 Jun 14 & 07:28 & 2454996.814   & $-$280.4     \\
CD$-$36~1052    & Magellan-Clay/MIKE      & 1200   & 2009 Jul 04 & 10:12 & 2455016.924   & $+$304.7     \\
CS 22876--040   & Magellan-Clay/MIKE      & 5800   & 2009 Jul 26 & 07:17 & 2455038.807   & $-$189.9     \\
\enddata
\end{deluxetable}

%% file: tab2.tex
\begin{deluxetable}{lcccccccccc}
\tablecaption{Observational Stellar Data
\label{rvtab}}
\tablewidth{0pt}
\tabletypesize{\scriptsize}
\tablehead{
\colhead{Star} &
\colhead{$\alpha$} &
\colhead{$\delta$} &
\colhead{$\langle$RV$\rangle$\tablenotemark{a}} & 
\colhead{Binary} &  
\colhead{Total exp.} &
\colhead{No.} & 
\colhead{S/N} & 
\colhead{S/N} & 
\colhead{S/N} & 
\colhead{S/N} \\ 
\colhead{} &
\colhead{(J2000)} &
\colhead{(J2000)} &
\colhead{(\kmsec)} &
\colhead{flag\tablenotemark{b}} &
\colhead{time (s)} &
\colhead{obs.} &
\colhead{3950\AA} &
\colhead{4550\AA} &
\colhead{5200\AA} &
\colhead{6750\AA} }
\startdata
BD$+$10~2495	& 12:59:20	& $+$09:14:36  & $+$258.3  & 3	& 9200	& 2	& 50	& 105	& 140	& 185   \\
BD$+$29~2356    & 13:01:52      & $+$29:11:18  & $-$207.3  & 1  & 5400  & 1     & 20    & 50    & 65    & 85    \\
BD$+$30~2611	& 15:06:54	& $+$30:00:37  & $-$280.3  & 1	& 7500	& 3	& 35	& 120	& 180	& 270   \\
CD$-$36~1052    & 02:47:38      & $-$36:06:24  & $+$304.7  & 1  & 1200  & 1     & 85    & 140   & 130   & 195   \\
CS 22876--040   & 00:04:52      & $-$34:13:37  & $-$189.9  & 0  & 5800  & 1     & 60    & 100   & 90    & 135   \\
CS 22948--093	& 21:50:32	& $-$41:07:49  & \nodata   & 2	& 10300 & 4     & 85    & 110   & 75    & 95    \\
CS 29513--031	& 23:25:11	& $-$39:59:29  & \nodata   & 2	& 7800  & 3     & 80    & 105   & 80    & 100   \\
CS 29513--032   & 23:22:19      & $-$39:44:25  & $-$215.6  & 1  & 4400  & 2     & 60    & 100   & 95    & 140   \\
\enddata 
\tablenotetext{a}{Mean or systemic Heliocentric radical velocity}
\tablenotetext{b}{Binary flags:
(0) unknown binary status;
(1) no RV variations detected in multiple epochs, either among our observations 
or in combination with previous studies;
(2) RV variations detected, suspected binary, no systemic velocity listed;
(3) spectroscopic binary confirmed by other studies, systemic velocity listed.
}
\end{deluxetable}

%% file: tab3.tex
\begin{deluxetable}{cl}
\tablecaption{References for Transition Probabilities or Oscillator Strengths
\label{loggftab}}
\tablewidth{0pt}
\tabletypesize{\scriptsize}
\tablehead{
\colhead{Species} &
\colhead{Reference(s) and Comments}}
\startdata
Li~\textsc{i}   & \citet{yan98} \\
C		& CH $A^2\Delta - X^2\Pi$ G-band linelist from B.\ Plez (private communication) \\
N               & CN $B^2\Sigma - X^2\Sigma$ (0,0) band from \citet{kurucz95} \\
O~\textsc{i}	& \citet{fuhr09} \\
Na~\textsc{i}	& \citet{fuhr09} \\
Mg~\textsc{i}	& \citet{aldenius07} for the Mg~b triplet; \citet{barklem05} \\
                & and references therein for all other transitions \\
Al~\textsc{i}	& \citet{fuhr09} \\
\enddata
\end{deluxetable}

%% file: tab4-stub.tex
\begin{deluxetable}{ccccc}
\tablecaption{Equivalent Width Measurements for All Stars
\label{ewtab}}
\tablewidth{0pt}
\tabletypesize{\scriptsize}
\tablehead{
\colhead{Wavelength (\AA)} &
\colhead{Species ID\tablenotemark{a}} &
\colhead{E.P.\ (eV)} &
\colhead{$\log (gf)$} &
\colhead{E.W.\ (m\AA)\tablenotemark{b}}}
\startdata
\multicolumn{5}{c}{                               BD$+$10~2495} \\ 
\hline       
   6707.80 &  3.0 & 0.00 &  0.17 &   synth \\
   6300.30 &  8.0 & 0.00 & -9.78 & \nodata \\
   7771.94 &  8.0 & 9.14 &  0.37 &     7.1 \\
   7774.17 &  8.0 & 9.14 &  0.22 &     7.5 \\
   7775.39 &  8.0 & 9.14 &  0.00 &     5.2 \\
\vdots & \vdots & \vdots & \vdots & \vdots \\
\enddata
\tablecomments{
The full table is available in machine-readable form in the electronic
edition of the journal;
only a small portion is shown here to present its general form and content.
}
\tablenotetext{a}{
The integer component of the species ID indicates the atomic number,
and the decimal component indicates the ionization state 
(0~$=$~neutral, 1~$=$~singly ionized).
}
\tablenotetext{b}{
``Synth'' indicates an abundance was derived from spectrum synthesis;
``limit'' indicates that an upper limit on the abundance was derived from 
the line.
}
\end{deluxetable}

%% file: tab5.tex
\begin{deluxetable}{cccccc}
\tablecaption{Model Atmosphere Parameters
\label{modeltab}}
\tablewidth{0pt}
\tablehead{
\colhead{Star} &
\colhead{T$_{\rm eff}$}&
\colhead{log $g$} & 
\colhead{$v_{t}$} &
\colhead{[M/H]} &
\colhead{Evolutionary} \\
\colhead{} &
\colhead{(K)} &
\colhead{} &
\colhead{(\kmsec)} &
\colhead{} &
\colhead{state} }
\startdata
         BD$+$10~2495 & 4710 &  1.30 &  1.55 & $-$2.45 & RGB  \\
         BD$+$29~2356 & 4760 &  1.60 &  1.45 & $-$1.55 & RGB  \\
         BD$+$30~2611 & 4330 &  0.60 &  1.85 & $-$1.45 & RGB  \\
         CD$-$36~1052 & 6070 &  2.30 &  3.20 & $-$1.65 & RHB  \\
        CS~22876--040 & 6280 &  3.70 &  1.60 & $-$2.30 & SGB  \\
        CS~22948--093 & 6480 &  4.00 &  1.15 & $-$3.35 & SGB  \\
        CS~29513--031 & 6650 &  3.70 &  0.80 & $-$2.55 & SGB  \\
        CS~29513--032 & 5810 &  3.30 &  1.45 & $-$2.10 & SGB  \\
            HD~119516 & 5290 &  1.10 &  2.25 & $-$2.15 & RHB  \\
            HD~128279 & 5050 &  2.35 &  1.50 & $-$2.45 & RGB  \\
            HD~175305 & 4770 &  1.80 &  1.25 & $-$1.60 & RGB  \\
            HD~237846 & 4600 &  1.00 &  1.50 & $-$3.20 & RGB  \\
\enddata
\end{deluxetable}

%% file: tab6.tex
\begin{deluxetable}{cccccc}
\tablecaption{Comparison of Temperature Estimates
\label{teffcomp}}
\tablewidth{0pt}
\tablehead{
\colhead{Star} &
\colhead{$T_{\rm eff}$\tablenotemark{a}} &
\colhead{$T_{\rm eff}(V-J)$\tablenotemark{b}} &
\colhead{$T_{\rm eff}(V-H)$\tablenotemark{b}} &
\colhead{$T_{\rm eff}(V-K)$\tablenotemark{b}} &
\colhead{$\Delta T_{\rm mean}$} \\
\colhead{} &
\colhead{(K)} &
\colhead{(K)} &
\colhead{(K)} &
\colhead{(K)} &
\colhead{(K)} }
\startdata
 BD$+$10~2495 & 4710 &  4946$\pm$119 & 4916$\pm$69 & 4964$\pm$74 & $-$232 \\
 BD$+$30~2611 & 4330 &  4258$\pm$209 & 4141$\pm$78 & 4244$\pm$40 & $+$116 \\
    HD~128279 & 5050 &  5274$\pm$157 & 5282$\pm$94 & 5305$\pm$82 & $-$237 \\
    HD~175305 & 4770 &  5063$\pm$152 & 5028$\pm$85 & 5042$\pm$66 & $-$274 \\
    HD~237846 & 4600 &  4968$\pm$123 & 4984$\pm$85 & 4985$\pm$65 & $-$379 \\ 
\enddata
\tablenotetext{a}{This study, spectroscopic estimate}
\tablenotetext{b}{\citet{alonso99a}}
\end{deluxetable}

%% file: tab7.tex
\begin{deluxetable}{ccccccccccccccc}
\tablecaption{Elemental Abundances I.\
\label{abundtab1}}
\tablewidth{0pt}
\tabletypesize{\scriptsize}
\tablehead{
\colhead{} &
\multicolumn{4}{c}{BD$+$10~2495} &
\colhead{} &
\multicolumn{4}{c}{BD$+$29~2356} &
\colhead{} &
\multicolumn{4}{c}{BD$+$30~2611} \\
\cline{2-5} \cline{7-10} \cline{12-15}
\colhead{Species} &
\colhead{log\,$\epsilon$(X)} &
\colhead{[X/Fe]} &
\colhead{$\sigma$} &
\colhead{N$_{\rm lines}$} &
\colhead{} &
\colhead{log($\epsilon$)} &
\colhead{[X/Fe]} &
\colhead{$\sigma$} &
\colhead{N$_{\rm lines}$} &
\colhead{} &
\colhead{log($\epsilon$)} &
\colhead{[X/Fe]} &
\colhead{$\sigma$} &
\colhead{N$_{\rm lines}$} \\
\colhead{} &
\colhead{(dex)} &
\colhead{(dex)} &
\colhead{(dex)} &
\colhead{} &
\colhead{} &
\colhead{(dex)} &
\colhead{(dex)} &
\colhead{(dex)} &
\colhead{} &
\colhead{} &
\colhead{(dex)} &
\colhead{(dex)} &
\colhead{(dex)} &
\colhead{} }
\startdata
Li I  &       0.57 &      -1.63 &    0.25 &       1  & &  $<$   0.70 &    \nodata & \nodata & \nodata  & & $<$  -0.01 &    \nodata & \nodata & \nodata  \\
C     &       5.92 &      -0.31 &    0.20 &       1  & &        6.71 &      -0.24 &    0.20 &       1  & &       6.32 &      -0.72 &    0.30 &       1  \\
N     &    \nodata &    \nodata & \nodata & \nodata  & &        6.57 &       0.36 &    0.20 &       1  & &    \nodata &    \nodata & \nodata & \nodata  \\
O  I  &       7.30 &       0.70 &    0.11 &       3  & &        8.04 &       0.72 &    0.25 &       1  & &       7.40 &      -0.01 &    0.30 &       1  \\
Na I  &       3.89 &      -0.11 &    0.25 &       1  & &        4.51 &      -0.21 &    0.15 &       2  & &       4.55 &      -0.26 &    0.15 &       2  \\
Mg I  &       5.78 &       0.53 &    0.15 &       4  & &        6.34 &       0.37 &    0.11 &       3  & &       6.37 &       0.31 &    0.16 &       3  \\
Al I  &       3.56 &      -0.58 &    0.25 &       1  & &     \nodata &    \nodata & \nodata & \nodata  & &    \nodata &    \nodata & \nodata & \nodata  \\
Si I  &       5.72 &       0.50 &    0.15 &       2  & &        6.10 &       0.16 &    0.23 &       4  & &       6.12 &       0.09 &    0.18 &       3  \\
\enddata
\end{deluxetable}

%% file: tab8.tex
\begin{deluxetable}{ccccccccccccccc}
\tablecaption{Elemental Abundances II.\
\label{abundtab2}}
\tablewidth{0pt}
\tabletypesize{\scriptsize}
\tablehead{
\colhead{} &
\multicolumn{4}{c}{CD$-$36~1052} &
\colhead{} &
\multicolumn{4}{c}{CS~22876--040} &
\colhead{} &
\multicolumn{4}{c}{CS~22948--093} \\
\cline{2-5} \cline{7-10} \cline{12-15}
\colhead{Species} &
\colhead{log\,$\epsilon$(X)} &
\colhead{[X/Fe]} &
\colhead{$\sigma$} &
\colhead{N$_{\rm lines}$} &
\colhead{} &
\colhead{log($\epsilon$)} &
\colhead{[X/Fe]} &
\colhead{$\sigma$} &
\colhead{N$_{\rm lines}$} &
\colhead{} &
\colhead{log($\epsilon$)} &
\colhead{[X/Fe]} &
\colhead{$\sigma$} &
\colhead{N$_{\rm lines}$} \\
\colhead{} &
\colhead{(dex)} &
\colhead{(dex)} &
\colhead{(dex)} &
\colhead{} &
\colhead{} &
\colhead{(dex)} &
\colhead{(dex)} &
\colhead{(dex)} &
\colhead{} &
\colhead{} &
\colhead{(dex)} &
\colhead{(dex)} &
\colhead{(dex)} &
\colhead{} }
\startdata
 Li I  & $<$   1.79 &    \nodata & \nodata & \nodata  & &       2.09 &      -0.11 &    0.25 &       1  & & $<$   2.35 &    \nodata & \nodata & \nodata  \\
 C     & $<$   7.05 & $<$   0.30 & \nodata & \nodata  & & $<$   6.79 & $<$   0.59 & \nodata & \nodata  & & $<$   6.90 & $<$   1.72 & \nodata & \nodata  \\
 N     & $<$   8.80 & $<$   2.79 & \nodata & \nodata  & & $<$   8.50 & $<$   3.04 & \nodata & \nodata  & & $<$   8.60 & $<$   4.16 & \nodata & \nodata  \\
 O  I  &       7.94 &       0.82 &    0.12 &       3  & &       7.20 &       0.63 &    0.10 &       3  & & $<$   7.23 & $<$   1.68 & \nodata & \nodata  \\
 Na I  &    \nodata &    \nodata & \nodata & \nodata  & &    \nodata &    \nodata & \nodata & \nodata  & & $<$   2.27 & $<$  -0.68 & \nodata & \nodata  \\
 Mg I  &       6.13 &       0.36 &    0.10 &       5  & &       5.53 &       0.31 &    0.10 &       4  & &       4.44 &       0.24 &    0.15 &       3  \\
 Al I  &       3.90 &      -0.76 &    0.15 &       2  & &       3.25 &      -0.86 &    0.25 &       1  & &       2.40 &      -0.69 &    0.15 &       2  \\
 Si I  &       6.06 &       0.32 &    0.25 &       1  & &       5.56 &       0.37 &    0.25 &       1  & &       4.15 &      -0.02 &    0.25 &       1  \\
\enddata
\end{deluxetable}

%% file: tab9.tex
\begin{deluxetable}{ccccccccccccccc}
\tablecaption{Elemental Abundances III.\
\label{abundtab3}}
\tablewidth{0pt}
\tabletypesize{\scriptsize}
\tablehead{
\colhead{} &
\multicolumn{4}{c}{CS~29513--031} &
\colhead{} &
\multicolumn{4}{c}{CS~29513--032} &
\colhead{} &
\multicolumn{4}{c}{HD~119516} \\
\cline{2-5} \cline{7-10} \cline{12-15}
\colhead{Species} &
\colhead{log\,$\epsilon$(X)} &
\colhead{[X/Fe]} &
\colhead{$\sigma$} &
\colhead{N$_{\rm lines}$} &
\colhead{} &
\colhead{log($\epsilon$)} &
\colhead{[X/Fe]} &
\colhead{$\sigma$} &
\colhead{N$_{\rm lines}$} &
\colhead{} &
\colhead{log($\epsilon$)} &
\colhead{[X/Fe]} &
\colhead{$\sigma$} &
\colhead{N$_{\rm lines}$} \\
\colhead{} &
\colhead{(dex)} &
\colhead{(dex)} &
\colhead{(dex)} &
\colhead{} &
\colhead{} &
\colhead{(dex)} &
\colhead{(dex)} &
\colhead{(dex)} &
\colhead{} &
\colhead{} &
\colhead{(dex)} &
\colhead{(dex)} &
\colhead{(dex)} &
\colhead{} }
\startdata
 Li I  & $<$   2.50 &    \nodata & \nodata & \nodata  & &       2.16 &      -0.04 &    0.25 &       1  & & $<$   1.14 &    \nodata & \nodata & \nodata  \\
 C     & $<$   7.10 & $<$   1.20 & \nodata & \nodata  & &       7.09 &       0.63 &    0.20 &       1  & & $<$   5.86 & $<$  -0.42 & \nodata & \nodata  \\
 N     & $<$   8.60 & $<$   3.19 & \nodata & \nodata  & & $<$   6.90 & $<$   1.04 & \nodata & \nodata  & & $<$   7.80 & $<$   2.26 & \nodata & \nodata  \\
 O  I  &       6.82 &       0.55 &    0.20 &       3  & &       7.48 &       0.65 &    0.10 &       3  & &       7.26 &       0.61 &    0.10 &       3  \\
 Na I  &       4.49 &       0.82 &    0.25 &       1  & &       4.69 &       0.46 &    0.25 &       1  & &       3.79 &      -0.26 &    0.25 &       1  \\
 Mg I  &       5.24 &       0.32 &    0.10 &       3  & &       6.00 &       0.52 &    0.14 &       5  & &       5.64 &       0.34 &    0.10 &       5  \\
 Al I  &       2.96 &      -0.85 &    0.30 &       1  & &       3.90 &      -0.49 &    0.22 &       2  & &       3.22 &      -0.97 &    0.16 &       2  \\
 Si I  &       4.92 &       0.03 &    0.25 &       1  & &       6.11 &       0.66 &    0.25 &       1  & &       5.81 &       0.54 &    0.25 &       1  \\
\enddata
\end{deluxetable}

%% file: tab10.tex
\begin{deluxetable}{ccccccccccccccc}
\tablecaption{Elemental Abundances IV.\
\label{abundtab4}}
\tablewidth{0pt}
\tabletypesize{\scriptsize}
\tablehead{
\colhead{} &
\multicolumn{4}{c}{HD~128279} &
\colhead{} &
\multicolumn{4}{c}{HD~175305} &
\colhead{} &
\multicolumn{4}{c}{HD~237846} \\
\cline{2-5} \cline{7-10} \cline{12-15}
\colhead{Species} &
\colhead{log\,$\epsilon$(X)} &
\colhead{[X/Fe]} &
\colhead{$\sigma$} &
\colhead{N$_{\rm lines}$} &
\colhead{} &
\colhead{log($\epsilon$)} &
\colhead{[X/Fe]} &
\colhead{$\sigma$} &
\colhead{N$_{\rm lines}$} &
\colhead{} &
\colhead{log($\epsilon$)} &
\colhead{[X/Fe]} &
\colhead{$\sigma$} &
\colhead{N$_{\rm lines}$} \\
\colhead{} &
\colhead{(dex)} &
\colhead{(dex)} &
\colhead{(dex)} &
\colhead{} &
\colhead{} &
\colhead{(dex)} &
\colhead{(dex)} &
\colhead{(dex)} &
\colhead{} &
\colhead{} &
\colhead{(dex)} &
\colhead{(dex)} &
\colhead{(dex)} &
\colhead{} }
\startdata
 Li I  &       0.91 &      -1.29 &    0.25 &       1  & &       0.73 &      -1.47 &    0.25 &       1  & &       0.55 &      -1.65 &    0.25 &       1  \\
 C     &       5.91 &      -0.12 &    0.15 &       1  & &       6.50 &      -0.31 &    0.15 &       1  & &       5.42 &       0.17 &    0.20 &       1  \\
 N     & $<$   5.70 & $<$   0.41 & \nodata & \nodata  & &    \nodata &    \nodata & \nodata & \nodata  & & $<$   5.70 & $<$   1.19 & \nodata & \nodata  \\
 O  I  &       7.32 &       0.92 &    0.10 &       3  & &       8.03 &       0.85 &    0.22 &       3  & &       7.08 &       1.46 &    0.25 &       1  \\
 Na I  &    \nodata &    \nodata & \nodata & \nodata  & &       4.44 &      -0.14 &    0.15 &       2  & &    \nodata &    \nodata & \nodata & \nodata  \\
 Mg I  &       5.58 &       0.53 &    0.10 &       4  & &       6.28 &       0.45 &    0.11 &       3  & &       4.79 &       0.52 &    0.10 &       2  \\
 Al I  &       3.40 &      -0.54 &    0.35 &       2  & &    \nodata &    \nodata & \nodata & \nodata  & &       2.58 &      -0.58 &    0.15 &       2  \\
 Si I  &       5.39 &       0.37 &    0.15 &       2  & &       6.02 &       0.22 &    0.20 &       4  & &    \nodata &    \nodata & \nodata & \nodata  \\
\enddata
\end{deluxetable}

%% file: tab11.tex
\begin{deluxetable}{cccccc}
\tablecaption{Star-to-star Dispersion in [X/Fe]
\label{dispersiontab}}
\tablewidth{0pt}
\tabletypesize{\scriptsize}
\tablehead{
\colhead{} &
\multicolumn{2}{c}{Stream Members} &
\colhead{} &
\multicolumn{2}{c}{\citet{cayrel04}} \\
\colhead{} &
\multicolumn{2}{c}{} &
\colhead{} &
\multicolumn{2}{c}{([Fe/H]~$ >-3.4$)} \\
\cline{2-3} \cline{5-6} 
\colhead{Ratio} &
\colhead{$\sigma_{\rm [X/Fe]}$} & 
\colhead{No.\ stars} &
\colhead{} &
\colhead{$\sigma_{\rm [X/Fe]}$} & 
\colhead{No.\ stars} }
\startdata
 ~[O~\textsc{i}/Fe]        & 0.36 & 10 & & 0.16 & 19 \\
 ~[Na~\textsc{i}/Fe]       & 0.07 &  5 & & 0.23 & 24 \\
 ~[Mg~\textsc{i}/Fe]       & 0.10 & 11 & & 0.13 & 26 \\
 ~[Al~\textsc{i}/Fe]       & 0.17 &  9 & & 0.18 & 26 \\
 ~[K~\textsc{i}/Fe]        & 0.21 &  6 & & 0.13 & 24 \\
 ~[Ca~\textsc{i}/Fe]       & 0.07 & 12 & & 0.08 & 26 \\
 ~[Sc~\textsc{ii}/Fe]      & 0.18 & 12 & & 0.10 & 26 \\
 ~[Ti~\textsc{ii}/Fe]      & 0.14 & 12 & & 0.10 & 26 \\
 ~[Cr~\textsc{i}/Fe]       & 0.05 & 12 & & 0.08 & 26 \\
 ~[Mn~\textsc{i}/Fe]       & 0.11 & 12 & & 0.10 & 26 \\
 ~[Co~\textsc{i}/Fe]       & 0.07 &  9 & & 0.11 & 26 \\
 ~[Ni~\textsc{i}/Fe]       & 0.06 & 12 & & 0.11 & 26 \\
 ~[Zn~\textsc{i}/Fe]       & 0.16 &  9 & & 0.09 & 26 \\
\enddata
\end{deluxetable}